\documentclass[prx,aps,twocolumn,footinbib,superscriptaddress,floatfix]{revtex4-2}
\usepackage{CJK}
\usepackage[colorlinks,bookmarks=false,citecolor=blue,linkcolor=red,urlcolor=blue]{hyperref}
\usepackage{color} 
\usepackage{graphicx}
\usepackage{float}
\usepackage{multirow}
\usepackage{amsmath}
\usepackage{bm}
\usepackage{amsfonts}
\usepackage{braket}
\usepackage{subfigure}
\usepackage{cleveref}

\usepackage{dcolumn}
\usepackage{amssymb}
\usepackage{microtype}
\usepackage{xfrac}
\usepackage{array}
\usepackage[dvipsnames]{xcolor}

\newcolumntype{P}[1]{>{\centering\arraybackslash}p{#1}}

\begin{document}

\title{Signatures of spinon dynamics and phase structure of\\ dipolar-octupolar quantum spin ices in  two-dimensional coherent spectroscopy }

\author{Mark Potts}
\affiliation{Max Planck Institute for the Physics of Complex Systems, N\"{o}thnitzer Str. 38, Dresden 01187, Germany}
\author{Roderich Moessner} 
\affiliation{Max Planck Institute for the Physics of Complex Systems, N\"{o}thnitzer Str. 38, Dresden 01187, Germany}
\author{Owen Benton} 
\affiliation{Max Planck Institute for the Physics of Complex Systems, N\"{o}thnitzer Str. 38, Dresden 01187, Germany}
\affiliation{School of Physical and Chemical Sciences, Queen Mary University of London, London, E1 4NS, United Kingdom}

\date{\today}

\begin{abstract}
We study how sharp signatures of fractionalization emerge in nonlinear spectroscopy experiments on spin liquids with separated energy scales. Our model is that of 
dipolar-octupolar rare earth pyrochlore materials, prime candidates for realizing quantum spin ice. This family of three dimensional quantum spin liquids exhibits fractionalization of spin degrees of freedom into spinons charged under an emergent $U(1)$ gauge field. We show that the technique of two dimensional coherent spectroscopy (2DCS) can identify clear signatures of fractionalised spinon dynamics in dipolar-octupolar quantum spin ices. However, at intermediate temperatures, spinon dynamics are heavily constrained in the presence of an incoherent spin background, leading to a {\it broad} 2DCS response. At lower temperatures, a sharp signal emerges as the system enters a  coherent spin liquid state. This lower temperature signal can in turn distinguish between zero-flux and $\pi$-flux forms of quantum spin ice. 
\end{abstract}

\maketitle

The emergence of fractional excitations \cite{rajaraman2001fractional,Moessner_Moore_2021} is a  feature of many exotic quantum states, including quantum spin liquids \cite{ANDERSON1973153,Moessner_2001,Balents10,Knolle19} and topologically ordered phases of matter \cite{Laughlin83, Wen02}. Establishing their signatures  is  key to the unambiguous identification of these phases in real materials. As fractional excitations can only be created in multiplets, kinematic considerations lead to broad linear response functions. These broad responses are difficult to differentiate from the response of more conventional excitations broadened due to (e.g.) disorder or finite lifetime effects.

The technique of two-dimensional coherent spectroscopy (2DCS) \cite{Wan19,Kuehn11,Woerner2013} has been proposed as a method that avoids this issue.
2DCS is a non-linear optical technique. Two weak electromagnetic pulses are applied at times $0$ and $\tau$, and the magnetization response measured at time $t+\tau$. The non-linear part of the response is then determined by subtracting off the contributions from both pulses alone. This non-linear response is a function of both $t$ and $\tau$, and its Fourier transform a function of $\omega_t$ and $\omega_{\tau}$. The broad continuum observed at linear response is found to produce a sharp extended signal along the $\omega_t=-\omega_{\tau}$ line of this Fourier transform, termed the `spinon-echo' or `rephasing' signal. Disorder effects and finite quasiparticle lifetimes introduce broadening transverse to the $\omega_t=-\omega_{\tau}$ line, which is crucially now distinguishable from the presence of a fractionalised continuum.

The technique has been investigated theoretically in various settings \cite{Wan19, Choi20, Hart20, Qiang23}, primarily in exactly solvable models such as the Kitaev honeycomb spin liquid, or certain one dimensional models. Here, we  discuss the application of 2DCS to quantum spin ice (QSI), a class of three dimensional spin liquids  on the pyrochlore lattice \cite{Hermele2004, Shannon2012, Benton2012, Savary2012, Gingras2014}. Several materials have been suggested as promising candidates for systems with QSI ground states \cite{Gingras2014, Kimura13, Sibille18, Gaudet2019, Sibille20,Poree22,poree2024}, although definitively establishing the presence of such a spin liquid in any specific material remains challenging. 

The most promising QSI candidate materials have been the rare earth pyrochlores, in particular the Ce based materials Ce$_2$Zr$_2$O$_7$, Ce$_2$Sn$_2$O$_7$, and Ce$_2$Hf$_2$O$_7$ \cite{Gaudet2019, Gao2019, Smith2023, Sibille20,Poree22,poree2024, Yahne2024}. These are all examples of so-called dipolar-octupolar pyrochlores due to the particular symmetry properties of their pseudospin-1/2 magnetic degrees of freedom. 

Here we discuss the capacity of 2DCS to identify fractionalised excitations in QSI. We present a theory for the 2DCS of dipolar-octupolar pyrochlores in two settings: (i) an intermediate temperature regime where spinon excitations are quantum coherent, but the initial state of the system can be taken to be an incoherent ensemble of classical spin ice ground states, which is treated using degenerate perturbation theory and exact diagonalization; (ii) the low energy regime where quantum coherence between spin ice states is important, which is treated using a gauge mean field theory (GMFT) \cite{Savary2012, Lee2012, Desrochers2023, Desrochers24}. 

\begin{figure}
    \centering
    \subfigure[]{\includegraphics[scale=0.14]{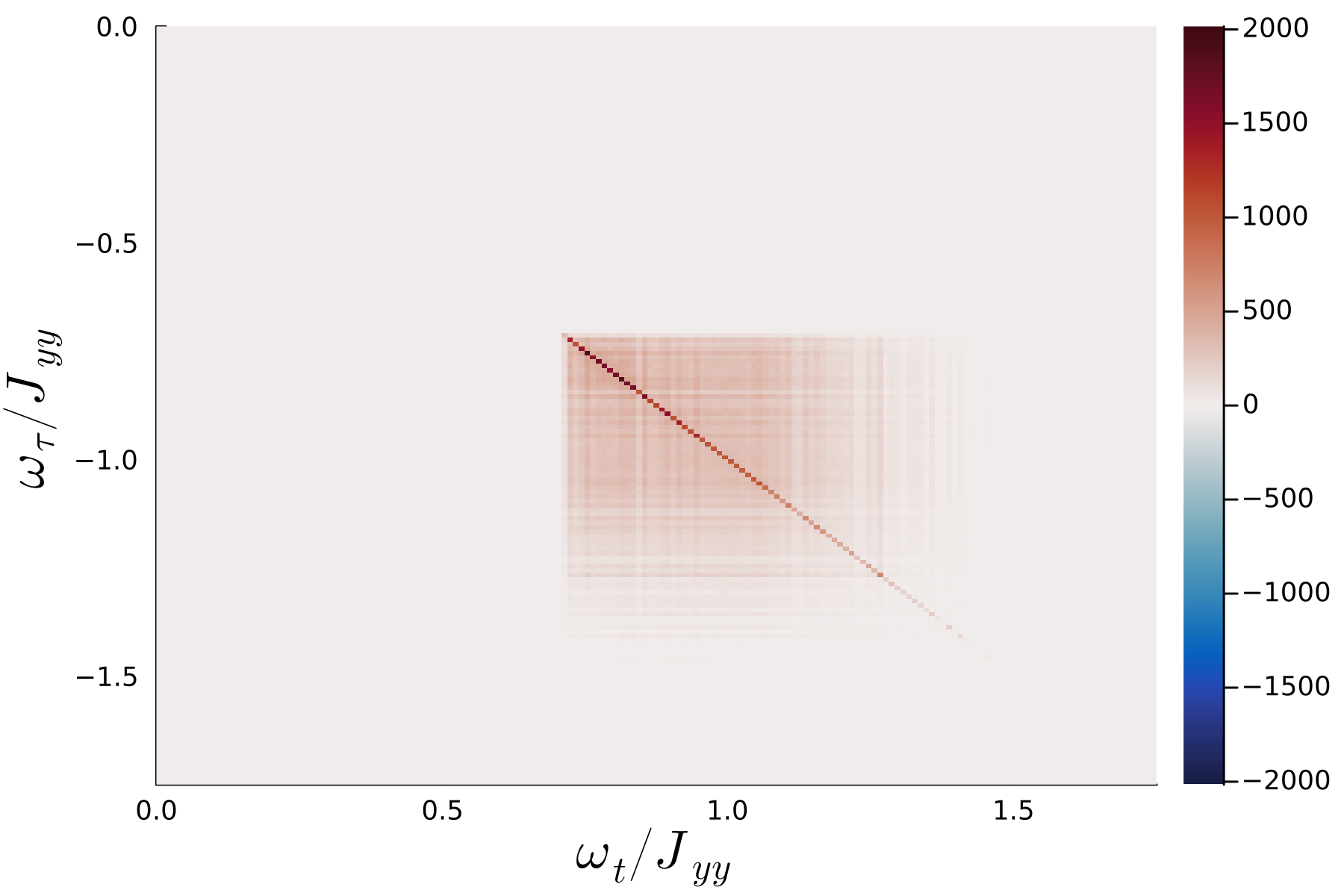}}
    \subfigure[]{\includegraphics[scale=0.14]{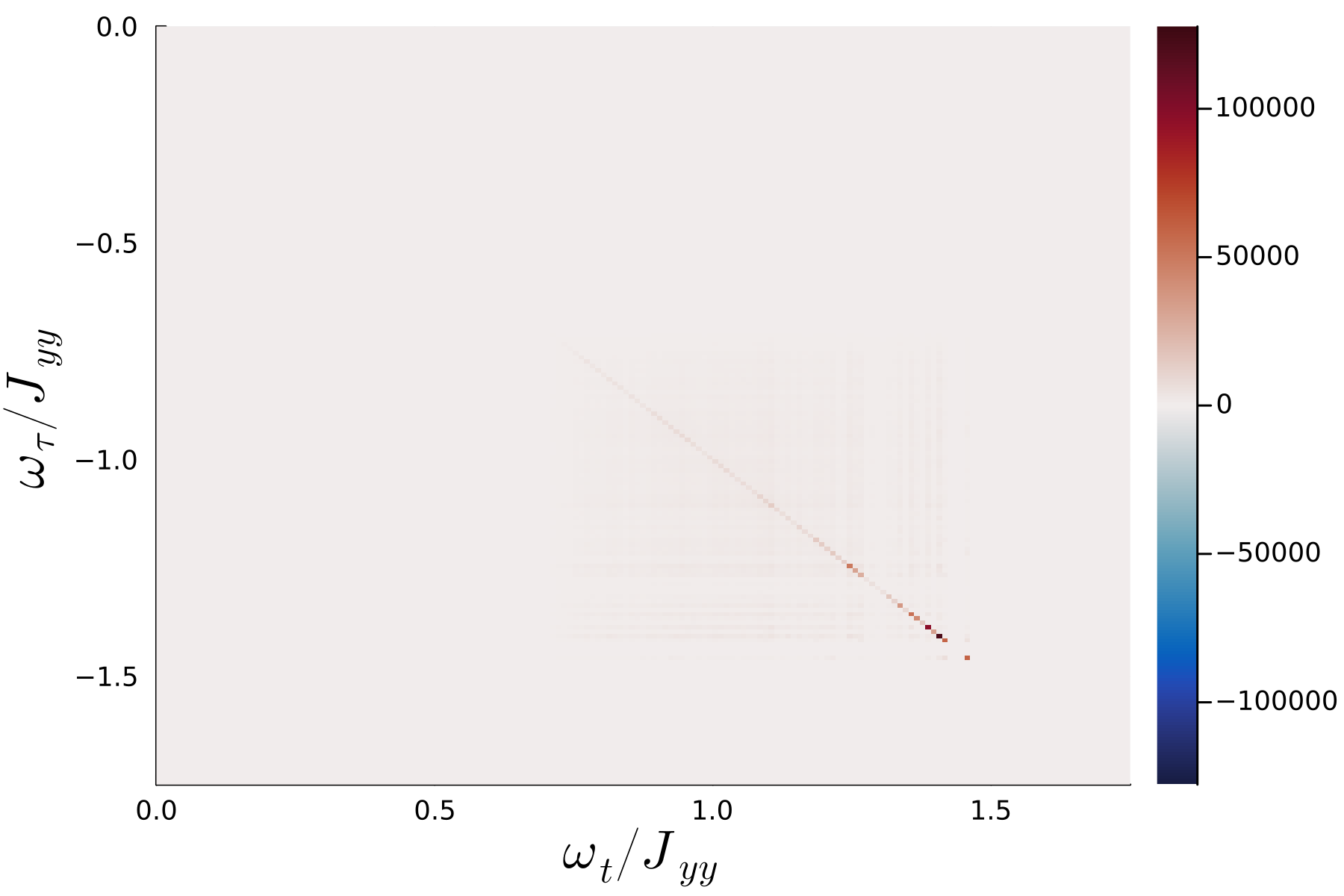}}
    \caption{Dimensionless 2DCS response $X(\omega_t,\omega_{\tau})(J_{yy}/2\hbar)^2$ of dipolar-octupolar QSI in the intermediate temperature regime with ((a)) and without ((b)) the geometrical factors $\mathbf{H}\cdot\mathbf{z}_i$ included in the field probe operator, as calculated using exact diagonalization on a finite cluster of 32 sites. A significant broad signal is observed if the correct geometrical coupling factors are included, due to the failure of the probe field operator to conserve the spinons' pseudo-momentum. By contrast, a much sharper rephasing streak is observed if the probe is modified to conserve pseudo-momentum by removing site dependent factors. The spectral weight of the response is shifted to the upper end of the spinon band, which can be understood in the Husimi picture in terms of an effective pseudo-momentum vertex factor. In both cases $J_{\pm}/J_{yy}=-0.05$ and the Lorentzian linewidth $\Gamma/J_{yy}=0.0005$.}
    \label{fig:2DCS_Husimi}
\end{figure}

\begin{figure}
    \centering
    \subfigure[]{\includegraphics[scale=0.23]{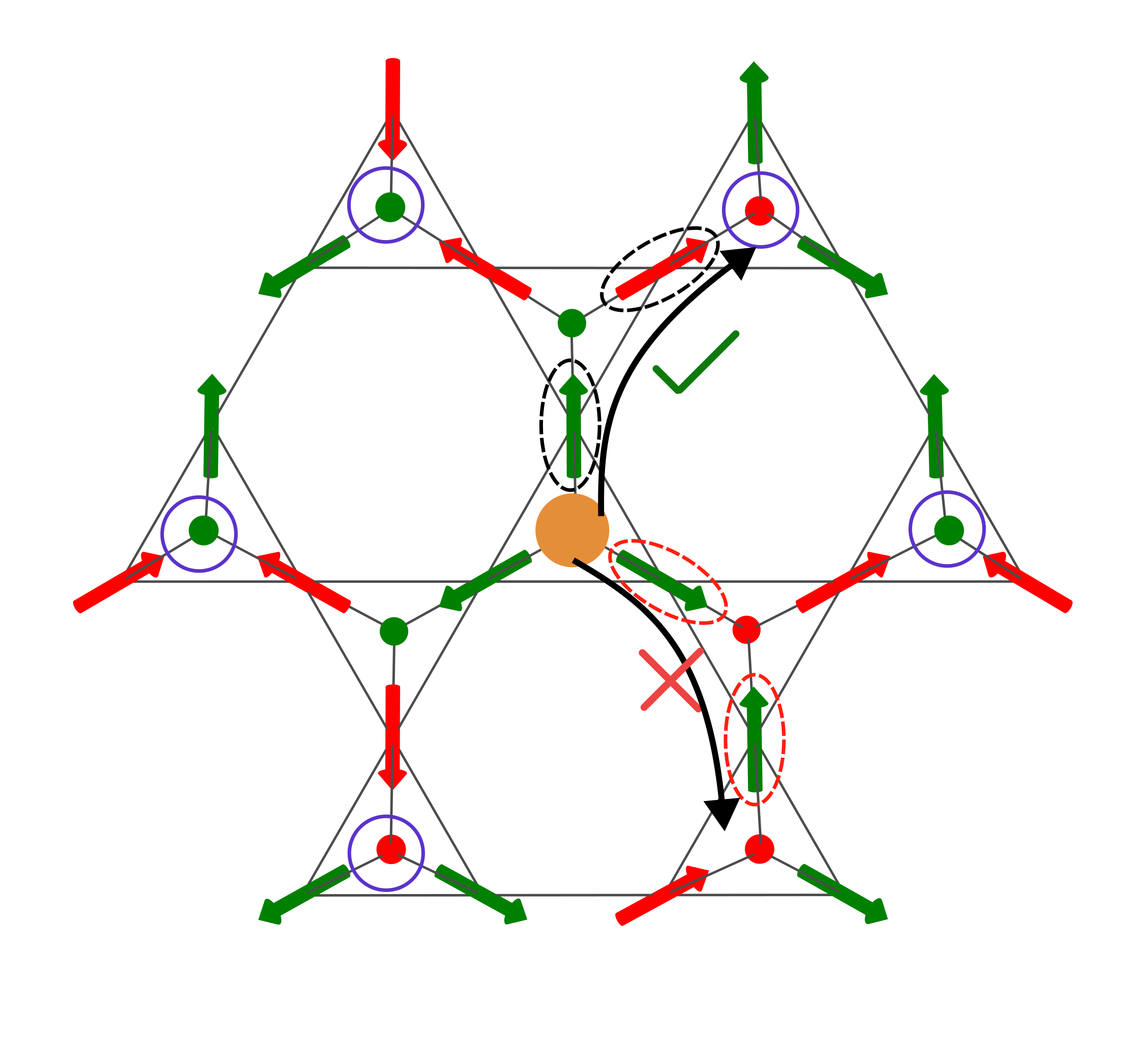}}
    \subfigure[]{\includegraphics[scale=0.23]{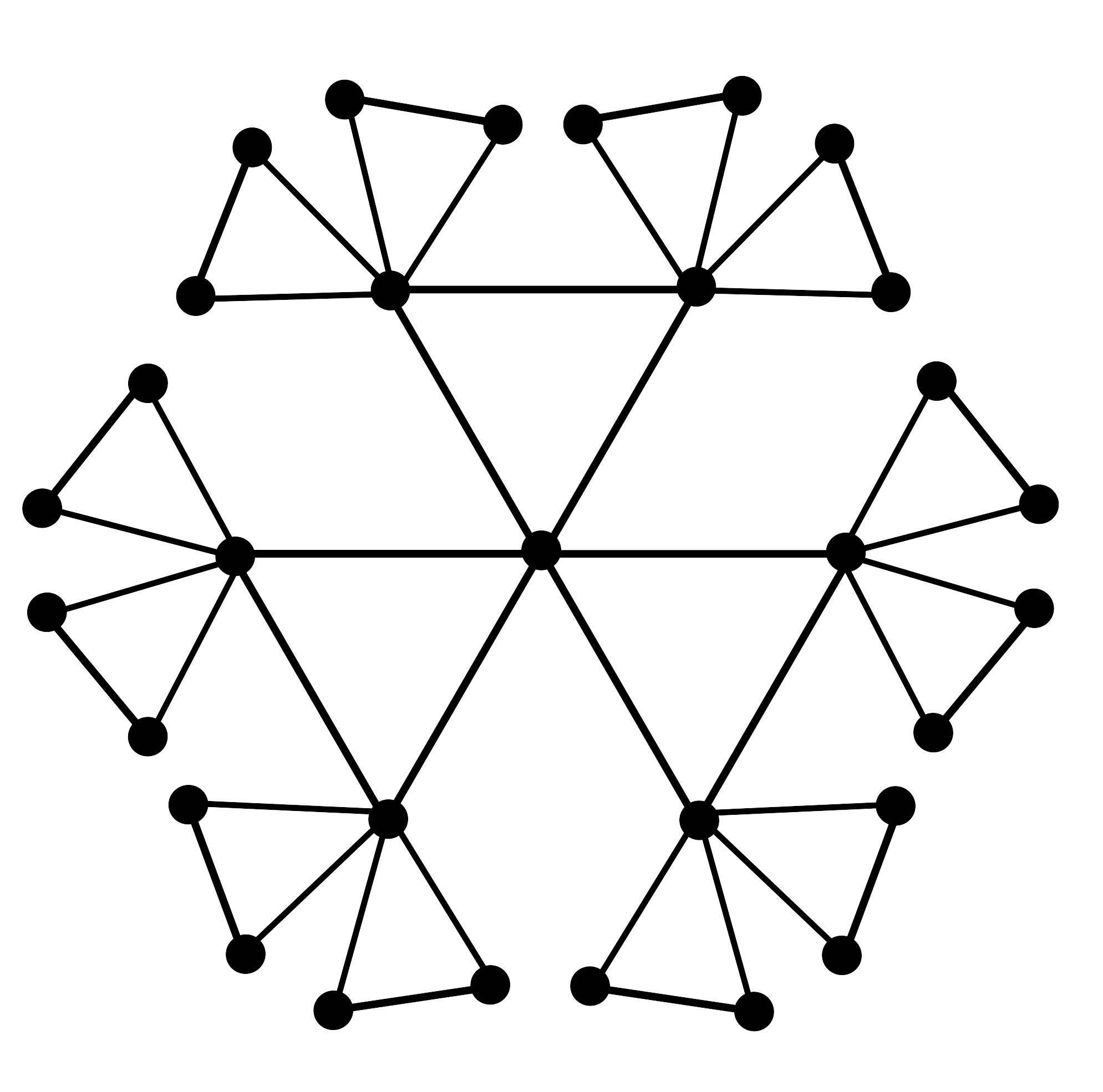}}
    \caption{(a) Section of the pyrochlore lattice with a single spinon indicated with a full orange circle. The perturbative dynamics allow spinon hops that flip two intermediate spins that are `head-to-tail'; only the sites indicated with a blue circle are accessible. (b): The Husimi cactus approximation for the many-body state space of a single spinon.}
    \label{fig:Hopping_diagram}
\end{figure}

We demonstrate that the 2DCS response produces a (perhaps surprising) broad response in the intermediate temperature regime [Fig. \ref{fig:2DCS_Husimi}]. We attribute this to the incoherent spin background and its constraints on spin spinon motion, which prevents these excitations from possessing well defined crystal momenta. We discuss how spinon motion can be approximated by hopping on a Husimi cactus [Fig. \ref{fig:Hopping_diagram}], where they acquire a one-dimensional pseudo-momentum quantum number which, due to the geometry of the lattice, is not preserved by a uniform probe field, hence the broad rephasing signal.

\begin{figure}
    \centering
    \subfigure[]{\includegraphics[scale=0.14]{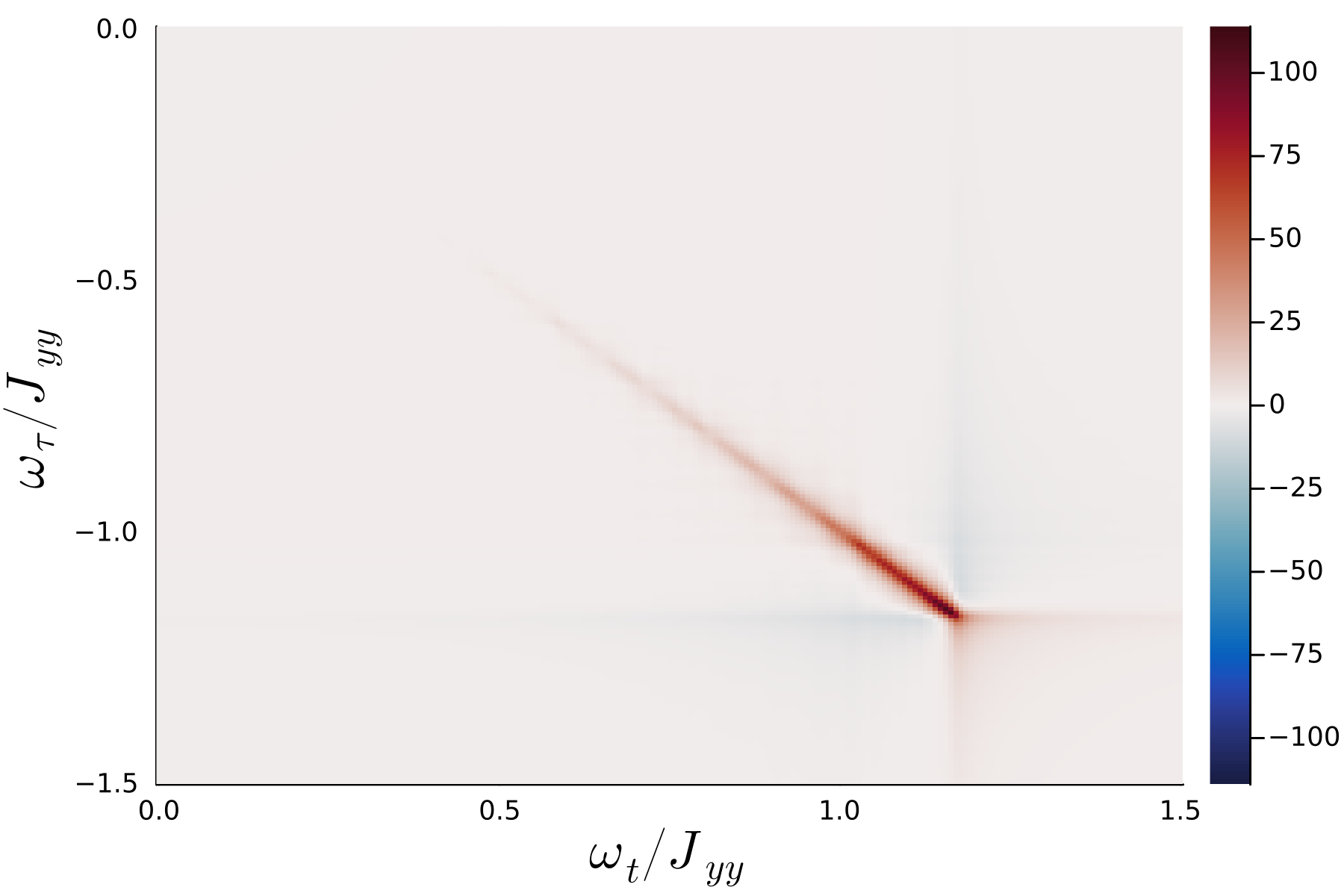}}
    \subfigure[]{\includegraphics[scale=0.14]{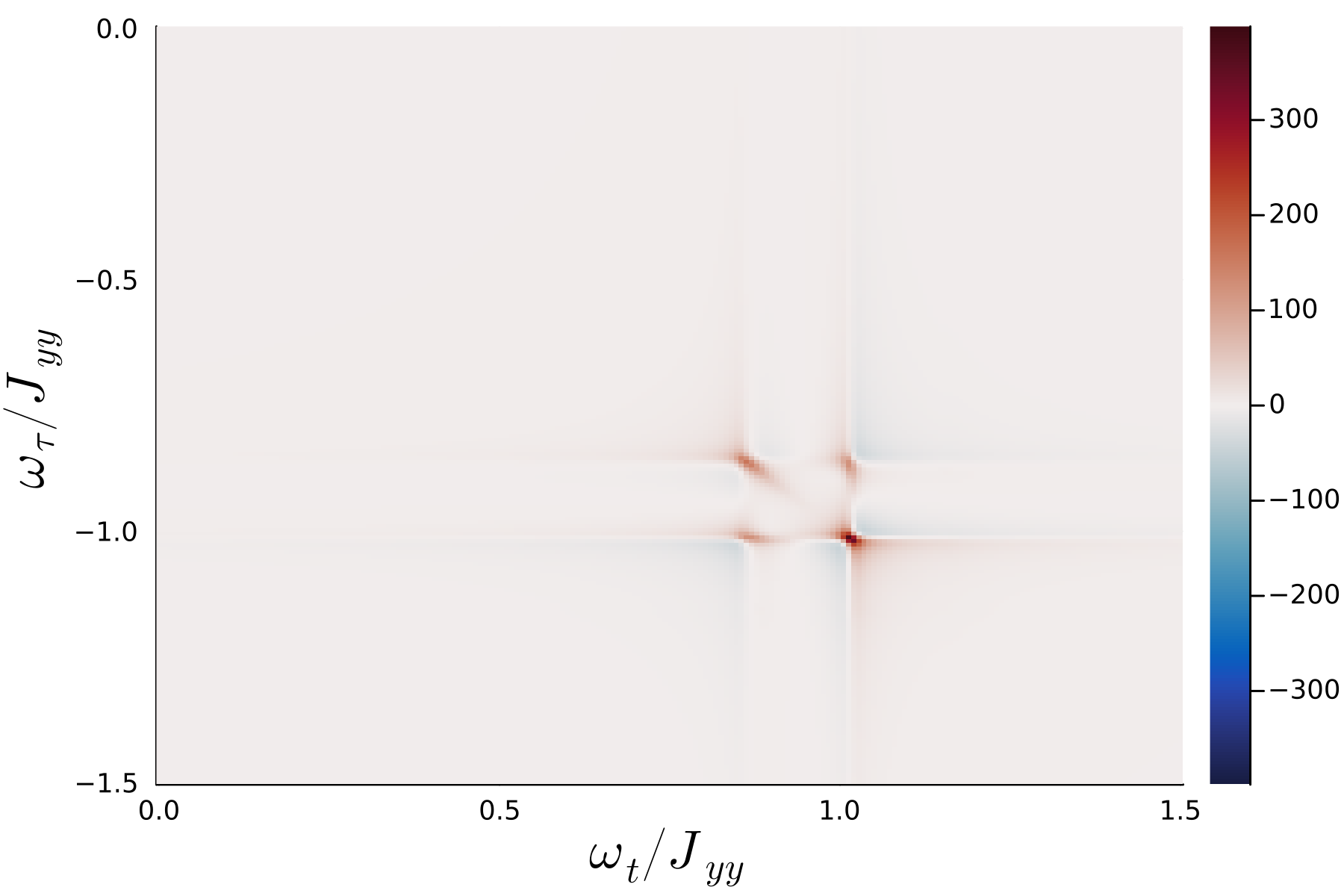}}
    \caption{Dimensionless 2DCS response $X(\omega_t,\omega_{\tau})(J_{yy}/\hbar)^2$ in the low temperature regime from GMFT. (a) Response in the $0$-flux phase (with $J_{\pm}/J_{yy}=0.04$); (b)  $\pi$-flux response (for $J_{\pm}/J_{yy}=-0.04)$. In the $0$-flux phase, a single extended streak is observed, whilst in the $\pi$-flux phase, the much narrower bandwidth of the two spinon bands, and transitions between them, result in multiple rephasing peaks, and additional peaks away from the main diagonal. Note that, if $\epsilon_{\pm}$ are the mean energies of each spinon band, the peaks in the $\pi$-flux response occur at frequencies $\omega_i=\epsilon_{-},$ $(\epsilon_{-}+\epsilon_{+})/2$. Matrix elements for processes where $\omega_i=\epsilon_{+}$ are heavily suppressed. In both cases the Lorentzian linewidth $\Gamma/J_{yy}=0.01$. Note that $\Gamma$ is here a factor 20 larger than in Fig. \ref{fig:2DCS_Husimi}, and thus the response is around an order of magnitude weaker.}
    \label{fig:2DCS_GMFT}
\end{figure}

At low temperatures the QSI ground state is a coherent superposition of classical ice states. Constraints on spinon motion are removed, and they recover well-defined crystal momenta. Sharp rephasing signals are obtained, contrasting with the results at intermediate temperatures [Fig. \ref{fig:2DCS_GMFT}]. Further, the 2DCS response clearly discriminates between the two possible QSI phases predicted for dipolar-ocutpolar pyrochlores in GMFT, the $0$-flux and $\pi$-flux $U(1)$ QSLs. In the $\pi$-flux phase, an increased unit cell size results in multiple spinon bands \cite{Desrochers24}, and transitions between them produce streaks away from the main rephasing diagonal.

Following a brief overview of dipolar-octupolar pyrochlores, we will discuss these items in turn. The most general nearest neighbor Hamiltonian for dipolar-octupolar pyrochlores is \cite{Rau19,Desrochers24}:
\begin{small}
\begin{equation}
    H=\sum_{\langle i,j \rangle} J_{yy} S^{y}_iS^{y}_j - J_{\pm}(S^{+}_iS^{-}_j+S^{-}_iS^{+}_j)+J_{\pm\pm}(S^{+}_iS^{+}_j+S^{-}_iS^{-}_j),\ \label{eq:General_Hamiltonian}
\end{equation}
\end{small}\noindent
with $\langle i,j \rangle$ denoting nearest neighbor sites on the pyrochlore lattice, and the $S^{\pm}$ operators being defined relative to the $y$ pseudospin direction. Motivated by the estimated parameter ranges for Ce pyrochlores \cite{Smith22, Sibille20,poree2024}, we assume that the dominant coupling is $J_{yy}>0$. The $S^{\mu}_j$ operators are pseudo-spin operators acting in the lowest energy Kramers doublet of the magnetic ions. Under the action of the point group of the lattice and time reversal, the $S^z$ and $S^x$ pseudo-spin operators transform as magnetic dipoles, whilst $S^y$ transforms as a magnetic octupole \cite{Rau19,Huang14}. For the basis chosen for Eq. (\ref{eq:General_Hamiltonian}),   $\mu_Bg_z \ \mathbf{H}\cdot\mathbf{z}_i \ (\cos\theta \ S^z_i + \sin\theta \ S^x_i)$ is the coupling to an external probe field $\mathbf{H}$. Here $\theta$ is a material dependent parameter, and the $\mathbf{z}_i$ are unit vectors along the lines connecting the centers of neighboring tetrahedra. These tetrahedra sit on two fcc sublattices, which we define as the `A' and `B' sublattices. This geometrical coupling factor between the external field and local degrees of freedom endows the probe field operator with some site dependence, which we find has a significant effect on the linear and non-linear responses of the system.

The rephasing signal that we calculate is only part of the full 2DCS response \cite{Wan19}. It is obtained from the imaginary part of the Fourier transform of the following non-linear susceptibility:
\begin{small}
\begin{align}
\chi^{(3)}(t,t,t+\tau) &= \left( \frac{g_z\mu_B}{\hbar}\right)^3\frac{g_z\mu_B}{V} X^{(3)}(t,t,t+\tau)  \\
X^{(3)}(t,t,t+\tau) &=\frac{i^3\theta(t)\theta(\tau)}{N}\langle [[[M(t+\tau),M(\tau)],M(\tau)],M(0)] \rangle, \label{eq:Non-linear_susceptibility}
\end{align}
\end{small}\noindent
with $M(t)=\sum_i\mathbf{H}\cdot\mathbf{z}_i/|\mathbf{H}| \ (\cos\theta \ S^z_i(t) + \sin\theta \ S^x_i(t))$  the time-dependent operator coupling the probe field and the pseudo-spin degrees of freedom, $V$  the volume of the unit cell, $g_z$ the dipolar g-factor, and $N$ the total number of spins. Polarization of the probe fields and the measurement axis are assumed to be identical, and  while $\chi^{(3)}(t,t,t+\tau)$ implicitly depends on the polarization, the response is found to be isotropic in all cases considered.
\medskip

%%%%%%%%%%%%%%%%%%%%%%%%%%%%%%%%%%%%%%%%

\textit{Intermediate temperature regime --}
In the limit of large $J_{yy}$, the Hamiltonian of Eq. (\ref{eq:General_Hamiltonian}) can be treated in degenerate perturbation theory.
The low energy states then obey a `2-in-2-out' ice rule on $S^y_i$ -- making the system an octupolar spin ice.
Tunneling between such configurations is of order $\sim|J_{\pm}|^3/J_{yy}^2$, while local violations of the ice rule (spinons) hop with matrix element $\sim J_{\pm}$.

At intermediate temperatures $|J_{\pm}|^3/J_{yy}^2, |J_{\pm\pm}|^3/J_{yy}^2 \ll T \ll |J_{\pm}|$ ($0.002 K < T < 0.08 K$ for one set of  exchange parameters for Ce$_2$Hf$_2$O$_7$ from   \cite{poree2024}), only the first order spinon hopping will be relevant, and the equilibrium state will be an incoherent ensemble of classical spin ice states. A schematic of the spinon hopping generated by the $J_{\pm}$ effective Hamiltonian is presented in Fig. \ref{fig:Hopping_diagram}. Note that this term in the Hamiltonian is symmetric under rotations in the $S^x$, $S^z$ plane, and so we can freely set $\theta$ to zero.

To obtain the 2DCS response in this regime, the $J_{\pm}$ perturbation  (second term in Eq. (\ref{eq:General_Hamiltonian})) is exactly diagonalized in the two and four spinon sectors on a 32-site pyrochlore cluster. 
The calculated 2DCS response in this regime is presented in the first panel of Fig. \ref{fig:2DCS_Husimi}. We introduce a small finite linewidth $\Gamma$ in our calculations to regulate delta-function divergences. To reduce the number of matrix elements that need to be calculated, $\Gamma$ is taken in the intermediate temperature regime to be relatively small. The smallness of $\Gamma$ leads to a relatively sharp onset of the response. Whilst a streak resembling the expected response for a system with fractionalised excitations is observed, it is accompanied by an unexpected broad region of non-zero intensity. 

We can understand this broad response by making use of an approximate analytic treatment for the spinon dynamics, wherein spinons hop on a Bethe lattice-like graph. 

To motivate this approach, we first note that the hopping generated by the perturbation Hamiltonian does not allow spinons to freely move across the lattice.
A spinon on the `A' fcc sublattice of tetrahedra will always remain on the `A' sublattice. Moreover, a spinon must hop by flipping two neighboring spins of opposite sign, or of opposite color in Fig. \ref{fig:Hopping_diagram} (into or out of an `A' tetrahedron), and must not generate any further spinons. This allows spinons to access only 6 of the 12 nearest neighbor fcc sites in a single hop, with the location of which depend on the local spin configuration.

To take these constraints on single spinon hopping into account, one should instead consider hopping on the full many-body state graph, tracking both the spinon location and the surrounding spin configuration \cite{Udagawa19}. As shown in Fig. \ref{fig:Hopping_diagram}b, taking just the motion of an isolated spinon, each state/vertex is connected to 6 further vertices. Spinons can hop in 3-cycles on the fcc lattice in such a way that both spinon location and the background spin configuration remain unchanged. Larger loops in state space result from the double traversal of loops on the fcc lattice with spins connected `head to tail'. Such loops do not appear regularly across the pyrochlore lattice, and thus one approximates the many-body graph for the motion of a single spinon by
a tree of triangles, with three triangles attached to each vertex.

In considering only the graph generated by single spinon hopping, interaction effects are ignored \cite{Udagawa19}. The approximations made to arrive at the Husimi cactus also prevent an exact mapping between spin configurations and Husimi states; only the local connectivity of the many-body state graph is maintained.

Within the Husimi cactus approximation, spinon dynamics undergo a dimensional reduction, and acquire a one-dimensional dispersion indexed by a pseudo-momentum $p$. This model for the spinon dynamics has been shown to successfully reproduce the linear responses of QSI within this temperature regime \cite{Udagawa19} when compared with exact diagonalization, \textit{a posteriori} justifying its use.

The Husimi perspective explains the broad features observed in the 2DCS response of Fig. \ref{fig:2DCS_Husimi}. The probe field operator $M(t)$ is proportional to $\sum_i\mathbf{H}\cdot\mathbf{z}_iS^z_i$ for dipolar-octupolar pyrochlores. The spatial dependence of this coupling prevents $M(t)$ from conserving spinon pseudo-momentum on the Husimi cactus. Without the conservation of pseudo-momentum, the `rephasing' that occurs in Eq. (\ref{eq:Non-linear_susceptibility}) to ensure a sharp streak response along $\omega_t=-\omega_{\tau}$ fails, and a broad response results. Further details concerning the representation of $M(t)$ in the Husimi cactus picture, and a demonstration that it does not conserve pseudo-momentum is presented in section \S II of the supplementary material \cite{supp}.

This explanation is verified by modifying the probe field operator $M(t)$ such that it has a definite pseudo-momentum. This is achieved by replacing the $\mathbf{H}\cdot\mathbf{z}_i$ coupling factors with a uniform constant H to remove its site dependence. Computed with the modified operator, the 2DCS response has a sharp rephasing streak, as shown in the second panel of Fig. \ref{fig:2DCS_Husimi}, supporting the conclusion that the broad features found with the physical probe field operator are a signature of the constrained spinon dynamics in the intermediate temperature regime.

\medskip

%%%%%%%%%%%%%%%%%%%%%%%%%%%%%%%%%%%%%%%%%%%

\begin{figure}
    \centering
    \includegraphics[scale=0.125]{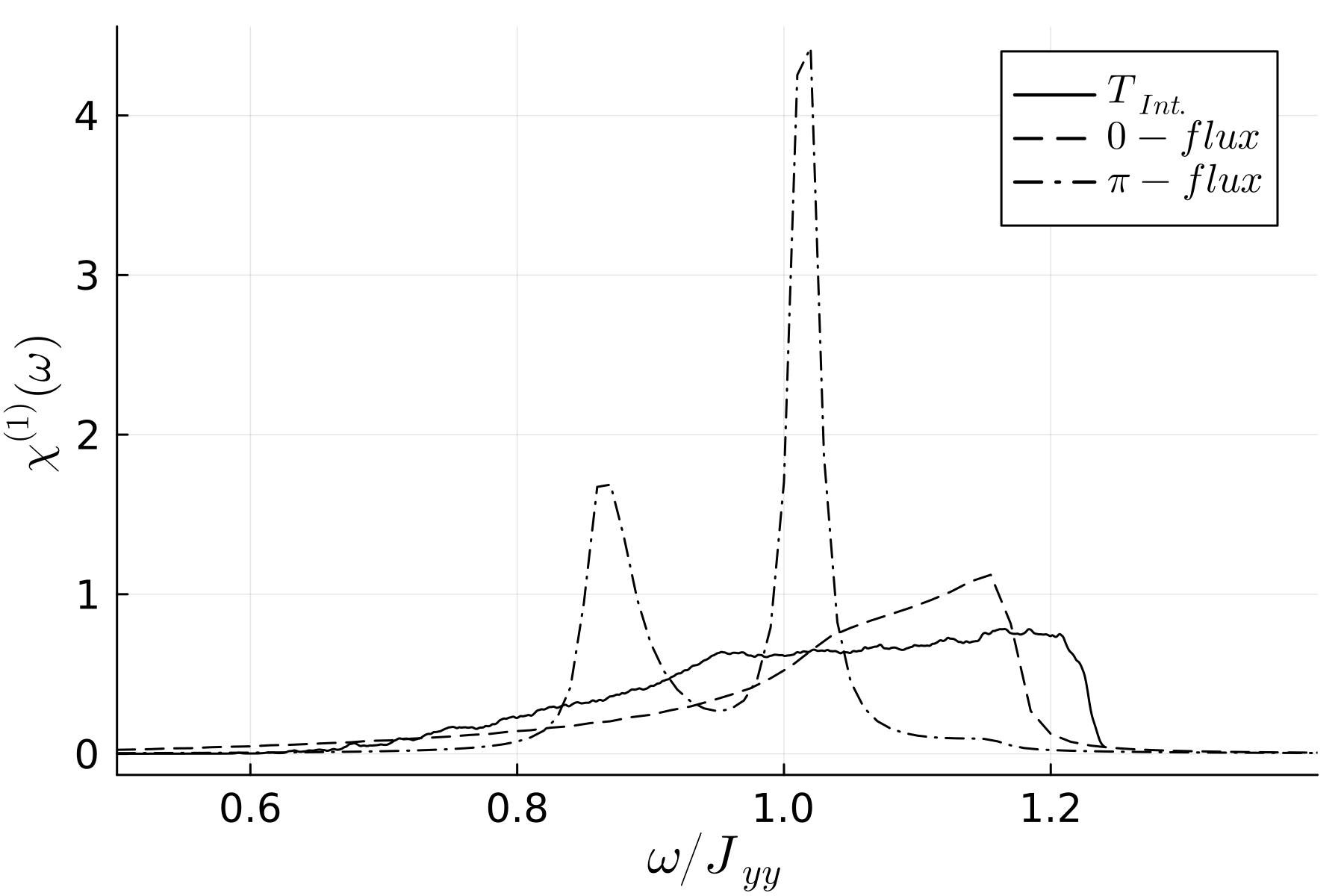}
    \caption{Dimensionless linear optical response functions $\chi(\omega) (g_z^2\mu_B^2/\hbar V)^{-1}(J_{yy}/\hbar)$ for the intermediate temperature regime ($T_{Int.}$), $0$-flux GMFT ground state, and $\pi$-flux GMFT ground state. For the intermediate temperature regime and $0$-flux response functions, $J_{\pm}/J_{yy}=0.04$, and for the $\pi$-flux  $J_{\pm}/J_{yy}=-0.04$. The two ground state phases remain distinguishable in the linear optical response, with the $\pi$-flux response possessing multiple clear maxima. However, whilst the broad features in the 2DCS response at intermediate temperatures clearly distinguish it from the low temperature 2DCS response, it is notable that the $T_{int}$ and $0$-flux linear responses are qualitatively very similar. For the ($T_{Int.}$) calculation, a Lorentzian linewidth $\Gamma/J_{yy}=0.002$ is used, whilst for the GMFT calculations $\Gamma/J_{yy}=0.01$.}
    \label{fig:Linear_responses}
\end{figure}

\textit{Low temperature regime --} The 2DCS response has broad features when quantum coherence between classical spin-ice states is suppressed by thermal fluctuations. By contrast, for $T\rightarrow0$, this anomalous broadening disappears, and sharp rephasing signals are recovered. In this limit, both gauge field and spinon degrees of freedom are quantum coherent. We analyze this regime by appealing to gauge mean field theory, making use of the formulation described in Ref. \cite{Desrochers24}. By the nature of mean field theories, this calculation is less controlled than the exact diagonalization analysis used in the intermediate temperature regime, however it has been shown to produce similar results to exact diagonalization and quantum Monte Carlo, where they can be compared \cite{Desrochers24}.

In GMFT, a gauge field $A_{i,j}$ and spinon field $\Phi_i$ are introduced: $S^{+}_{i,j}=\Phi^{\dagger}e^{iA_{i,j}}\Phi_j/2$, where $i$ and $j$ label neighboring tetrahedra on the `A' and `B' sublattices respectively.  $S^y_{i,j}$ maps to an emergent electric field $E_{i,j}$ obeying an emergent Gauss' law. Technical details can be found in Ref. \cite{Desrochers24}, where it is also demonstrated that there exist two possible mean field solutions for gapped spinon effective actions consistent with the gauge, time-reversal and lattice symmetries. These are the $0$-flux and $\pi$-flux $U(1)$ dipolar-ocutpolar spin ice phases. In both phases, spinons on the `A' and `B' sublattices decouple, and the terms generated by the $J_{\pm\pm}$ term in Eq. (\ref{eq:General_Hamiltonian}) vanish in the mean field, again allowing us to freely set the mixing angle $\theta$ to zero. In the 0-flux phase, the `A' and `B' sublattice spinons both have a single dispersive band, whilst in the $\pi$-flux phase, the threading of $\pi$-flux through each hexagonal plaquette of the pyrochlore lattice leads to an enlargement of the unit cell, resulting in two doubly degenerate bands per sublattice. 

The two-band dispersion of spinons in the $\pi$-flux phase has already been identified as a signature distinguishing it from the $0$-flux regime in linear response \cite{Lee2012,desrochers2024finite}, and we find that spinon transitions between these two bands produce features in the 2DCS response that clearly distinguish the two phases. Figure \ref{fig:2DCS_GMFT} presents the 2DCS responses for both phases calculated within GMFT using the physical field coupling operator $M(t)=\mathbf{H}\cdot\mathbf{z}_i/|\mathbf{H}| \ S^z_i(t)$. For the 0-flux phase, a single rephasing streak is obtained, whilst the $\pi$-flux response resembles a cluster of isolated peaks. The reason for this stark contrast is two-fold. First, each of the two distinct spinon bands in the $\pi$-flux phase are relatively flat leading to a density of states with two well defined maxima, and hence relatively narrow bandwidth optical responses. Second, the probe operator has non-zero matrix elements to excite a spinon from one band to another, and these inter-band transitions result in the additional peaks off of the main rephasing diagonal. 
\medskip

%%%%%%%%%%%%%%%%%%%%%%%%%%%%%%%%%%%

\textit{Discussion --} The 2DCS responses of spinons in dipolar-octupolar QSI are found to be  sensitive to both temperature regime and phase of the system. Sharp responses reflecting the presence of fractionalized spinon excitations are predicted for the lowest $T$, and these responses are demonstrated to clearly distinguish between the spinon dynamics found in the 0-flux and $\pi$-flux phases of dipolar-octupolar spin ice. Conversely, in the intermediate temperature regime, a broad non-linear response is observed indicative of heavily constrained spinon motion in the presence of an incoherent spin background.

Whilst this letter has focused on the example of dipolar-octupolar pyrochlores, we hypothesize that the appearance of broad signatures in 2DCS where fractionalised excitations move across an incoherent spin background may occur more generally in similar frustrated models, presenting an obstacle to observing sharp signatures of fractionalisation at all but the lowest temperatures. On the other hand, by measuring the 2DCS signal as a function of temperature one could use the crossover between these regimes to demonstrate the onset of a quantum coherent low-temperature state. The GMFT calculation however does not include the effects of quantum fluctuations in the gauge field or interactions between the spinons and photon modes, which will be necessary to understand the cross-over between these two regimes.

A natural question is then whether the clear distinctions between the spinon dynamical regimes observable in 2DCS can be seen  directly in the {\it linear} optical response to a uniform magnetic field, which is more simply accessed experimentally. In Fig. \ref{fig:Linear_responses} we present the linear responses for all phases and regimes considered above. The linear response varies between each regime, and in particular the existence of multiple maxima in the $\pi$-flux response distinguishes it from the $T=0$ $0$-flux response, as already noted in Ref. \cite{Desrochers24}. However, the stark difference between the behavior of the response in the intermediate temperature regime and the low temperature regime is entirely absent; the intermediate temperature response and the low temperature $0$-flux response are not qualitatively different, despite very different underlying spinon dynamics. By contrast, the existence of broad features in the intermediate temperature 2DCS response clearly distinguishes this regime from the low-temperature regime.

To summarize, we find that 2DCS is a suitable experimental setting to probe the dynamics of fractionalised excitations in candidate quantum spin ice materials, providing clear signatures distinguishing the regime where fractionalised quasiparticles move ballistically on the pyrochlore lattice at the lowest temperatures from that where their motion is heavily constrained by an incoherent spin background. Further, the appearance of response streaks away from the main rephasing diagonal in 2DCS offers a direct signature of the transition of quasiparticles between bands, which for dipolar-octupolar quantum spin ices allows one to distinguish experimentally between a $0$-flux or $\pi$-flux QSI ground states. We believe that this further demonstrates the potential utility of 2DCS as a probe of the dynamics of fractionalised excitations in quantum spin ice candidate materials, and other systems potentially hosting exotic quantum phases of matter.
\medskip

\begin{acknowledgments}
We thank Benedikt Placke and Yuan Wan for useful discussions. This work was supported in part by
the Deutsche Forschungsgemeinschaft under Grant No. SFB
1143 (Project-ID No. 247310070) and by
the Deutsche Forschungsgemeinschaft  under cluster of excellence
ct.qmat (EXC 2147, Project-ID No. 390858490).
\end{acknowledgments}

\bibliography{Newbib2.bib}

%apsrev4-2.bst 2019-01-14 (MD) hand-edited version of apsrev4-1.bst
%Control: key (0)
%Control: author (8) initials jnrlst
%Control: editor formatted (1) identically to author
%Control: production of article title (0) allowed
%Control: page (0) single
%Control: year (1) truncated
%Control: production of eprint (0) enabled
\begin{thebibliography}{39}%
\makeatletter
\providecommand \@ifxundefined [1]{%
 \@ifx{#1\undefined}
}%
\providecommand \@ifnum [1]{%
 \ifnum #1\expandafter \@firstoftwo
 \else \expandafter \@secondoftwo
 \fi
}%
\providecommand \@ifx [1]{%
 \ifx #1\expandafter \@firstoftwo
 \else \expandafter \@secondoftwo
 \fi
}%
\providecommand \natexlab [1]{#1}%
\providecommand \enquote  [1]{``#1''}%
\providecommand \bibnamefont  [1]{#1}%
\providecommand \bibfnamefont [1]{#1}%
\providecommand \citenamefont [1]{#1}%
\providecommand \href@noop [0]{\@secondoftwo}%
\providecommand \href [0]{\begingroup \@sanitize@url \@href}%
\providecommand \@href[1]{\@@startlink{#1}\@@href}%
\providecommand \@@href[1]{\endgroup#1\@@endlink}%
\providecommand \@sanitize@url [0]{\catcode `\\12\catcode `\$12\catcode
  `\&12\catcode `\#12\catcode `\^12\catcode `\_12\catcode `\%12\relax}%
\providecommand \@@startlink[1]{}%
\providecommand \@@endlink[0]{}%
\providecommand \url  [0]{\begingroup\@sanitize@url \@url }%
\providecommand \@url [1]{\endgroup\@href {#1}{\urlprefix }}%
\providecommand \urlprefix  [0]{URL }%
\providecommand \Eprint [0]{\href }%
\providecommand \doibase [0]{https://doi.org/}%
\providecommand \selectlanguage [0]{\@gobble}%
\providecommand \bibinfo  [0]{\@secondoftwo}%
\providecommand \bibfield  [0]{\@secondoftwo}%
\providecommand \translation [1]{[#1]}%
\providecommand \BibitemOpen [0]{}%
\providecommand \bibitemStop [0]{}%
\providecommand \bibitemNoStop [0]{.\EOS\space}%
\providecommand \EOS [0]{\spacefactor3000\relax}%
\providecommand \BibitemShut  [1]{\csname bibitem#1\endcsname}%
\let\auto@bib@innerbib\@empty
%</preamble>
\bibitem [{\citenamefont {Rajaraman}(2001)}]{rajaraman2001fractional}%
  \BibitemOpen
  \bibfield  {author} {\bibinfo {author} {\bibfnamefont {R.}~\bibnamefont
  {Rajaraman}},\ }\href@noop {} {\bibinfo {title} {Fractional charge}}
  (\bibinfo {year} {2001}),\ \Eprint {https://arxiv.org/abs/cond-mat/0103366}
  {arXiv:cond-mat/0103366 [cond-mat.mes-hall]} \BibitemShut {NoStop}%
\bibitem [{\citenamefont {Moessner}\ and\ \citenamefont
  {Moore}(2021)}]{Moessner_Moore_2021}%
  \BibitemOpen
  \bibfield  {author} {\bibinfo {author} {\bibfnamefont {R.}~\bibnamefont
  {Moessner}}\ and\ \bibinfo {author} {\bibfnamefont {J.~E.}\ \bibnamefont
  {Moore}},\ }\href@noop {} {\emph {\bibinfo {title} {Topological Phases of
  Matter}}}\ (\bibinfo  {publisher} {Cambridge University Press},\ \bibinfo
  {year} {2021})\BibitemShut {NoStop}%
\bibitem [{\citenamefont {Anderson}(1973)}]{ANDERSON1973153}%
  \BibitemOpen
  \bibfield  {author} {\bibinfo {author} {\bibfnamefont {P.}~\bibnamefont
  {Anderson}},\ }\bibfield  {title} {\bibinfo {title} {Resonating valence
  bonds: A new kind of insulator?},\ }\href
  {https://doi.org/https://doi.org/10.1016/0025-5408(73)90167-0} {\bibfield
  {journal} {\bibinfo  {journal} {Materials Research Bulletin}\ }\textbf
  {\bibinfo {volume} {8}},\ \bibinfo {pages} {153} (\bibinfo {year}
  {1973})}\BibitemShut {NoStop}%
\bibitem [{\citenamefont {Moessner}\ and\ \citenamefont
  {Sondhi}(2001)}]{Moessner_2001}%
  \BibitemOpen
  \bibfield  {author} {\bibinfo {author} {\bibfnamefont {R.}~\bibnamefont
  {Moessner}}\ and\ \bibinfo {author} {\bibfnamefont {S.~L.}\ \bibnamefont
  {Sondhi}},\ }\bibfield  {title} {\bibinfo {title} {Resonating valence bond
  phase in the triangular lattice quantum dimer model},\ }\href
  {https://doi.org/10.1103/PhysRevLett.86.1881} {\bibfield  {journal} {\bibinfo
   {journal} {Phys. Rev. Lett.}\ }\textbf {\bibinfo {volume} {86}},\ \bibinfo
  {pages} {1881} (\bibinfo {year} {2001})}\BibitemShut {NoStop}%
\bibitem [{\citenamefont {Balents}(2010)}]{Balents10}%
  \BibitemOpen
  \bibfield  {author} {\bibinfo {author} {\bibfnamefont {L.}~\bibnamefont
  {Balents}},\ }\bibfield  {title} {\bibinfo {title} {Spin liquids in
  frustrated magnets},\ }\href {https://doi.org/10.1038/nature08917} {\bibfield
   {journal} {\bibinfo  {journal} {Nature}\ }\textbf {\bibinfo {volume}
  {464}},\ \bibinfo {pages} {199–208} (\bibinfo {year} {2010})}\BibitemShut
  {NoStop}%
\bibitem [{\citenamefont {Knolle}\ and\ \citenamefont
  {Moessner}(2019)}]{Knolle19}%
  \BibitemOpen
  \bibfield  {author} {\bibinfo {author} {\bibfnamefont {J.}~\bibnamefont
  {Knolle}}\ and\ \bibinfo {author} {\bibfnamefont {R.}~\bibnamefont
  {Moessner}},\ }\bibfield  {title} {\bibinfo {title} {A field guide to spin
  liquids},\ }\href {https://doi.org/10.1146/annurev-conmatphys-031218-013401}
  {\bibfield  {journal} {\bibinfo  {journal} {Annual Review of Condensed Matter
  Physics}\ }\textbf {\bibinfo {volume} {10}},\ \bibinfo {pages} {451}
  (\bibinfo {year} {2019})},\ \Eprint
  {https://arxiv.org/abs/https://doi.org/10.1146/annurev-conmatphys-031218-013401}
  {https://doi.org/10.1146/annurev-conmatphys-031218-013401} \BibitemShut
  {NoStop}%
\bibitem [{\citenamefont {Laughlin}(1983)}]{Laughlin83}%
  \BibitemOpen
  \bibfield  {author} {\bibinfo {author} {\bibfnamefont {R.~B.}\ \bibnamefont
  {Laughlin}},\ }\bibfield  {title} {\bibinfo {title} {Anomalous quantum hall
  effect: An incompressible quantum fluid with fractionally charged
  excitations},\ }\href {https://doi.org/10.1103/PhysRevLett.50.1395}
  {\bibfield  {journal} {\bibinfo  {journal} {Phys. Rev. Lett.}\ }\textbf
  {\bibinfo {volume} {50}},\ \bibinfo {pages} {1395} (\bibinfo {year}
  {1983})}\BibitemShut {NoStop}%
\bibitem [{\citenamefont {Wen}(2002)}]{Wen02}%
  \BibitemOpen
  \bibfield  {author} {\bibinfo {author} {\bibfnamefont {X.-G.}\ \bibnamefont
  {Wen}},\ }\bibfield  {title} {\bibinfo {title} {Quantum orders and symmetric
  spin liquids},\ }\href {https://doi.org/10.1103/PhysRevB.65.165113}
  {\bibfield  {journal} {\bibinfo  {journal} {Phys. Rev. B}\ }\textbf {\bibinfo
  {volume} {65}},\ \bibinfo {pages} {165113} (\bibinfo {year}
  {2002})}\BibitemShut {NoStop}%
\bibitem [{\citenamefont {Wan}\ and\ \citenamefont {Armitage}(2019)}]{Wan19}%
  \BibitemOpen
  \bibfield  {author} {\bibinfo {author} {\bibfnamefont {Y.}~\bibnamefont
  {Wan}}\ and\ \bibinfo {author} {\bibfnamefont {N.~P.}\ \bibnamefont
  {Armitage}},\ }\bibfield  {title} {\bibinfo {title} {Resolving continua of
  fractional excitations by spinon echo in thz 2d coherent spectroscopy},\
  }\href {https://doi.org/10.1103/PhysRevLett.122.257401} {\bibfield  {journal}
  {\bibinfo  {journal} {Phys. Rev. Lett.}\ }\textbf {\bibinfo {volume} {122}},\
  \bibinfo {pages} {257401} (\bibinfo {year} {2019})}\BibitemShut {NoStop}%
\bibitem [{\citenamefont {Kuehn}\ \emph {et~al.}(2011)\citenamefont {Kuehn},
  \citenamefont {Reimann}, \citenamefont {Woerner}, \citenamefont {Elsaesser},\
  and\ \citenamefont {Hey}}]{Kuehn11}%
  \BibitemOpen
  \bibfield  {author} {\bibinfo {author} {\bibfnamefont {W.}~\bibnamefont
  {Kuehn}}, \bibinfo {author} {\bibfnamefont {K.}~\bibnamefont {Reimann}},
  \bibinfo {author} {\bibfnamefont {M.}~\bibnamefont {Woerner}}, \bibinfo
  {author} {\bibfnamefont {T.}~\bibnamefont {Elsaesser}},\ and\ \bibinfo
  {author} {\bibfnamefont {R.}~\bibnamefont {Hey}},\ }\bibfield  {title}
  {\bibinfo {title} {Two-dimensional terahertz correlation spectra of
  electronic excitations in semiconductor quantum wells},\ }\href
  {https://doi.org/10.1021/jp1099046} {\bibfield  {journal} {\bibinfo
  {journal} {The Journal of Physical Chemistry B}\ }\textbf {\bibinfo {volume}
  {115}},\ \bibinfo {pages} {5448} (\bibinfo {year} {2011})},\ \bibinfo {note}
  {pMID: 21171588},\ \Eprint
  {https://arxiv.org/abs/https://doi.org/10.1021/jp1099046}
  {https://doi.org/10.1021/jp1099046} \BibitemShut {NoStop}%
\bibitem [{\citenamefont {Woerner}\ \emph {et~al.}(2013)\citenamefont
  {Woerner}, \citenamefont {Kuehn}, \citenamefont {Bowlan}, \citenamefont
  {Reimann},\ and\ \citenamefont {Elsaesser}}]{Woerner2013}%
  \BibitemOpen
  \bibfield  {author} {\bibinfo {author} {\bibfnamefont {M.}~\bibnamefont
  {Woerner}}, \bibinfo {author} {\bibfnamefont {W.}~\bibnamefont {Kuehn}},
  \bibinfo {author} {\bibfnamefont {P.}~\bibnamefont {Bowlan}}, \bibinfo
  {author} {\bibfnamefont {K.}~\bibnamefont {Reimann}},\ and\ \bibinfo {author}
  {\bibfnamefont {T.}~\bibnamefont {Elsaesser}},\ }\bibfield  {title} {\bibinfo
  {title} {Ultrafast two-dimensional terahertz spectroscopy of elementary
  excitations in solids},\ }\href
  {https://doi.org/10.1088/1367-2630/15/2/025039} {\bibfield  {journal}
  {\bibinfo  {journal} {New Journal of Physics}\ }\textbf {\bibinfo {volume}
  {15}},\ \bibinfo {pages} {025039} (\bibinfo {year} {2013})}\BibitemShut
  {NoStop}%
\bibitem [{\citenamefont {Choi}\ \emph {et~al.}(2020)\citenamefont {Choi},
  \citenamefont {Lee},\ and\ \citenamefont {Kim}}]{Choi20}%
  \BibitemOpen
  \bibfield  {author} {\bibinfo {author} {\bibfnamefont {W.}~\bibnamefont
  {Choi}}, \bibinfo {author} {\bibfnamefont {K.~H.}\ \bibnamefont {Lee}},\ and\
  \bibinfo {author} {\bibfnamefont {Y.~B.}\ \bibnamefont {Kim}},\ }\bibfield
  {title} {\bibinfo {title} {Theory of two-dimensional nonlinear spectroscopy
  for the kitaev spin liquid},\ }\href
  {https://doi.org/10.1103/PhysRevLett.124.117205} {\bibfield  {journal}
  {\bibinfo  {journal} {Phys. Rev. Lett.}\ }\textbf {\bibinfo {volume} {124}},\
  \bibinfo {pages} {117205} (\bibinfo {year} {2020})}\BibitemShut {NoStop}%
\bibitem [{\citenamefont {Hart}\ and\ \citenamefont
  {Nandkishore}(2023)}]{Hart20}%
  \BibitemOpen
  \bibfield  {author} {\bibinfo {author} {\bibfnamefont {O.}~\bibnamefont
  {Hart}}\ and\ \bibinfo {author} {\bibfnamefont {R.}~\bibnamefont
  {Nandkishore}},\ }\bibfield  {title} {\bibinfo {title} {Extracting spinon
  self-energies from two-dimensional coherent spectroscopy},\ }\href
  {https://doi.org/10.1103/PhysRevB.107.205143} {\bibfield  {journal} {\bibinfo
   {journal} {Phys. Rev. B}\ }\textbf {\bibinfo {volume} {107}},\ \bibinfo
  {pages} {205143} (\bibinfo {year} {2023})}\BibitemShut {NoStop}%
\bibitem [{\citenamefont {Qiang}\ \emph {et~al.}()\citenamefont {Qiang},
  \citenamefont {Quito}, \citenamefont {Trevisan},\ and\ \citenamefont
  {Orth}}]{Qiang23}%
  \BibitemOpen
  \bibfield  {author} {\bibinfo {author} {\bibfnamefont {Y.}~\bibnamefont
  {Qiang}}, \bibinfo {author} {\bibfnamefont {V.~L.}\ \bibnamefont {Quito}},
  \bibinfo {author} {\bibfnamefont {T.~V.}\ \bibnamefont {Trevisan}},\ and\
  \bibinfo {author} {\bibfnamefont {P.~P.}\ \bibnamefont {Orth}},\ }\bibfield
  {title} {\bibinfo {title} {Probing majorana wavefunctions in kitaev honeycomb
  spin liquids with second-order two-dimensional spectroscopy},\ }\href
  {https://arxiv.org/abs/2301.11243} {\bibinfo  {journal} {arXiv:2301.11243}\
  }\BibitemShut {NoStop}%
\bibitem [{\citenamefont {Hermele}\ \emph {et~al.}(2004)\citenamefont
  {Hermele}, \citenamefont {Fisher},\ and\ \citenamefont
  {Balents}}]{Hermele2004}%
  \BibitemOpen
\bibfield  {journal} {  }\bibfield  {author} {\bibinfo {author} {\bibfnamefont
  {M.}~\bibnamefont {Hermele}}, \bibinfo {author} {\bibfnamefont {M.~P.~A.}\
  \bibnamefont {Fisher}},\ and\ \bibinfo {author} {\bibfnamefont
  {L.}~\bibnamefont {Balents}},\ }\bibfield  {title} {\bibinfo {title}
  {Pyrochlore photons: The $u(1)$ spin liquid in a $s=\frac{1}{2}$
  three-dimensional frustrated magnet},\ }\href
  {https://doi.org/10.1103/PhysRevB.69.064404} {\bibfield  {journal} {\bibinfo
  {journal} {Phys. Rev. B}\ }\textbf {\bibinfo {volume} {69}},\ \bibinfo
  {pages} {064404} (\bibinfo {year} {2004})}\BibitemShut {NoStop}%
\bibitem [{\citenamefont {Shannon}\ \emph {et~al.}(2012)\citenamefont
  {Shannon}, \citenamefont {Sikora}, \citenamefont {Pollmann}, \citenamefont
  {Penc},\ and\ \citenamefont {Fulde}}]{Shannon2012}%
  \BibitemOpen
  \bibfield  {author} {\bibinfo {author} {\bibfnamefont {N.}~\bibnamefont
  {Shannon}}, \bibinfo {author} {\bibfnamefont {O.}~\bibnamefont {Sikora}},
  \bibinfo {author} {\bibfnamefont {F.}~\bibnamefont {Pollmann}}, \bibinfo
  {author} {\bibfnamefont {K.}~\bibnamefont {Penc}},\ and\ \bibinfo {author}
  {\bibfnamefont {P.}~\bibnamefont {Fulde}},\ }\bibfield  {title} {\bibinfo
  {title} {Quantum ice: A quantum monte carlo study},\ }\href
  {https://doi.org/10.1103/PhysRevLett.108.067204} {\bibfield  {journal}
  {\bibinfo  {journal} {Phys. Rev. Lett.}\ }\textbf {\bibinfo {volume} {108}},\
  \bibinfo {pages} {067204} (\bibinfo {year} {2012})}\BibitemShut {NoStop}%
\bibitem [{\citenamefont {Benton}\ \emph {et~al.}(2012)\citenamefont {Benton},
  \citenamefont {Sikora},\ and\ \citenamefont {Shannon}}]{Benton2012}%
  \BibitemOpen
  \bibfield  {author} {\bibinfo {author} {\bibfnamefont {O.}~\bibnamefont
  {Benton}}, \bibinfo {author} {\bibfnamefont {O.}~\bibnamefont {Sikora}},\
  and\ \bibinfo {author} {\bibfnamefont {N.}~\bibnamefont {Shannon}},\
  }\bibfield  {title} {\bibinfo {title} {Seeing the light: Experimental
  signatures of emergent electromagnetism in a quantum spin ice},\ }\href
  {https://doi.org/10.1103/PhysRevB.86.075154} {\bibfield  {journal} {\bibinfo
  {journal} {Phys. Rev. B}\ }\textbf {\bibinfo {volume} {86}},\ \bibinfo
  {pages} {075154} (\bibinfo {year} {2012})}\BibitemShut {NoStop}%
\bibitem [{\citenamefont {Savary}\ and\ \citenamefont
  {Balents}(2012)}]{Savary2012}%
  \BibitemOpen
  \bibfield  {author} {\bibinfo {author} {\bibfnamefont {L.}~\bibnamefont
  {Savary}}\ and\ \bibinfo {author} {\bibfnamefont {L.}~\bibnamefont
  {Balents}},\ }\bibfield  {title} {\bibinfo {title} {Coulombic quantum liquids
  in spin-$1/2$ pyrochlores},\ }\href
  {https://doi.org/10.1103/PhysRevLett.108.037202} {\bibfield  {journal}
  {\bibinfo  {journal} {Phys. Rev. Lett.}\ }\textbf {\bibinfo {volume} {108}},\
  \bibinfo {pages} {037202} (\bibinfo {year} {2012})}\BibitemShut {NoStop}%
\bibitem [{\citenamefont {Gingras}\ and\ \citenamefont
  {McClarty}(2014)}]{Gingras2014}%
  \BibitemOpen
  \bibfield  {author} {\bibinfo {author} {\bibfnamefont {M.~J.~P.}\
  \bibnamefont {Gingras}}\ and\ \bibinfo {author} {\bibfnamefont {P.~A.}\
  \bibnamefont {McClarty}},\ }\bibfield  {title} {\bibinfo {title} {Quantum
  spin ice: a search for gapless quantum spin liquids in pyrochlore magnets},\
  }\href {https://doi.org/10.1088/0034-4885/77/5/056501} {\bibfield  {journal}
  {\bibinfo  {journal} {Reports on Progress in Physics}\ }\textbf {\bibinfo
  {volume} {77}},\ \bibinfo {pages} {056501} (\bibinfo {year}
  {2014})}\BibitemShut {NoStop}%
\bibitem [{\citenamefont {Kimura}\ \emph {et~al.}(2013)\citenamefont {Kimura},
  \citenamefont {Nakatsuji}, \citenamefont {Wen}, \citenamefont {Broholm},
  \citenamefont {Stone}, \citenamefont {Nishibori},\ and\ \citenamefont
  {Sawa}}]{Kimura13}%
  \BibitemOpen
  \bibfield  {author} {\bibinfo {author} {\bibfnamefont {K.}~\bibnamefont
  {Kimura}}, \bibinfo {author} {\bibfnamefont {S.}~\bibnamefont {Nakatsuji}},
  \bibinfo {author} {\bibfnamefont {J.-J.}\ \bibnamefont {Wen}}, \bibinfo
  {author} {\bibfnamefont {C.}~\bibnamefont {Broholm}}, \bibinfo {author}
  {\bibfnamefont {M.~B.}\ \bibnamefont {Stone}}, \bibinfo {author}
  {\bibfnamefont {E.}~\bibnamefont {Nishibori}},\ and\ \bibinfo {author}
  {\bibfnamefont {H.}~\bibnamefont {Sawa}},\ }\bibfield  {title} {\bibinfo
  {title} {Quantum fluctuations in spin-ice-like pr$_2$zr$_2$o$_7$},\ }\href
  {https://www.nature.com/articles/ncomms2914} {\bibfield  {journal} {\bibinfo
  {journal} {Nature Commun.}\ }\textbf {\bibinfo {volume} {4}},\ \bibinfo
  {pages} {1934} (\bibinfo {year} {2013})}\BibitemShut {NoStop}%
\bibitem [{\citenamefont {Sibille}\ \emph {et~al.}(2018)\citenamefont
  {Sibille}, \citenamefont {Gauthier}, \citenamefont {Yan}, \citenamefont
  {Ciomaga~Hatnean}, \citenamefont {Olliver}, \citenamefont {Winn},
  \citenamefont {Filges}, \citenamefont {Balakrishnan}, \citenamefont
  {Kenzelmann}, \citenamefont {Shannon},\ and\ \citenamefont
  {Fennell}}]{Sibille18}%
  \BibitemOpen
  \bibfield  {author} {\bibinfo {author} {\bibfnamefont {R.}~\bibnamefont
  {Sibille}}, \bibinfo {author} {\bibfnamefont {N.}~\bibnamefont {Gauthier}},
  \bibinfo {author} {\bibfnamefont {H.}~\bibnamefont {Yan}}, \bibinfo {author}
  {\bibfnamefont {M.}~\bibnamefont {Ciomaga~Hatnean}}, \bibinfo {author}
  {\bibfnamefont {J.}~\bibnamefont {Olliver}}, \bibinfo {author} {\bibfnamefont
  {B.}~\bibnamefont {Winn}}, \bibinfo {author} {\bibfnamefont {U.}~\bibnamefont
  {Filges}}, \bibinfo {author} {\bibfnamefont {G.}~\bibnamefont
  {Balakrishnan}}, \bibinfo {author} {\bibfnamefont {M.}~\bibnamefont
  {Kenzelmann}}, \bibinfo {author} {\bibfnamefont {N.}~\bibnamefont
  {Shannon}},\ and\ \bibinfo {author} {\bibfnamefont {T.}~\bibnamefont
  {Fennell}},\ }\bibfield  {title} {\bibinfo {title} {Experimental signatures
  of emergent quantum electrodynamics in pr$_2$hf$_2$o$_7$},\ }\href
  {https://www.nature.com/articles/s41567-018-0116-x} {\bibfield  {journal}
  {\bibinfo  {journal} {Nat. Phys.}\ }\textbf {\bibinfo {volume} {14}},\
  \bibinfo {pages} {711} (\bibinfo {year} {2018})}\BibitemShut {NoStop}%
\bibitem [{\citenamefont {Gaudet}\ \emph {et~al.}(2019)\citenamefont {Gaudet},
  \citenamefont {Smith}, \citenamefont {Dudemaine}, \citenamefont {Beare},
  \citenamefont {Buhariwalla}, \citenamefont {Butch}, \citenamefont {Stone},
  \citenamefont {Kolesnikov}, \citenamefont {Xu}, \citenamefont {Yahne},
  \citenamefont {Ross}, \citenamefont {Marjerrison}, \citenamefont {Garrett},
  \citenamefont {Luke}, \citenamefont {Bianchi},\ and\ \citenamefont
  {Gaulin}}]{Gaudet2019}%
  \BibitemOpen
  \bibfield  {author} {\bibinfo {author} {\bibfnamefont {J.}~\bibnamefont
  {Gaudet}}, \bibinfo {author} {\bibfnamefont {E.~M.}\ \bibnamefont {Smith}},
  \bibinfo {author} {\bibfnamefont {J.}~\bibnamefont {Dudemaine}}, \bibinfo
  {author} {\bibfnamefont {J.}~\bibnamefont {Beare}}, \bibinfo {author}
  {\bibfnamefont {C.~R.~C.}\ \bibnamefont {Buhariwalla}}, \bibinfo {author}
  {\bibfnamefont {N.~P.}\ \bibnamefont {Butch}}, \bibinfo {author}
  {\bibfnamefont {M.~B.}\ \bibnamefont {Stone}}, \bibinfo {author}
  {\bibfnamefont {A.~I.}\ \bibnamefont {Kolesnikov}}, \bibinfo {author}
  {\bibfnamefont {G.}~\bibnamefont {Xu}}, \bibinfo {author} {\bibfnamefont
  {D.~R.}\ \bibnamefont {Yahne}}, \bibinfo {author} {\bibfnamefont {K.~A.}\
  \bibnamefont {Ross}}, \bibinfo {author} {\bibfnamefont {C.~A.}\ \bibnamefont
  {Marjerrison}}, \bibinfo {author} {\bibfnamefont {J.~D.}\ \bibnamefont
  {Garrett}}, \bibinfo {author} {\bibfnamefont {G.~M.}\ \bibnamefont {Luke}},
  \bibinfo {author} {\bibfnamefont {A.~D.}\ \bibnamefont {Bianchi}},\ and\
  \bibinfo {author} {\bibfnamefont {B.~D.}\ \bibnamefont {Gaulin}},\ }\bibfield
   {title} {\bibinfo {title} {Quantum spin ice dynamics in the dipole-octupole
  pyrochlore magnet ${\mathrm{ce}}_{2}{\mathrm{zr}}_{2}{\mathrm{o}}_{7}$},\
  }\href {https://doi.org/10.1103/PhysRevLett.122.187201} {\bibfield  {journal}
  {\bibinfo  {journal} {Phys. Rev. Lett.}\ }\textbf {\bibinfo {volume} {122}},\
  \bibinfo {pages} {187201} (\bibinfo {year} {2019})}\BibitemShut {NoStop}%
\bibitem [{\citenamefont {Sibille}\ \emph {et~al.}(2020)\citenamefont
  {Sibille}, \citenamefont {Gauthier},\ and\ \citenamefont
  {Lhotel}}]{Sibille20}%
  \BibitemOpen
  \bibfield  {author} {\bibinfo {author} {\bibfnamefont {R.}~\bibnamefont
  {Sibille}}, \bibinfo {author} {\bibfnamefont {N.}~\bibnamefont {Gauthier}},\
  and\ \bibinfo {author} {\bibfnamefont {E.~e.~a.}\ \bibnamefont {Lhotel}},\
  }\bibfield  {title} {\bibinfo {title} {A quantum liquid of magnetic octupoles
  on the pyrochlore lattice},\ }\href
  {https://doi.org/10.1038/s41567-020-0827-7} {\bibfield  {journal} {\bibinfo
  {journal} {Nat. Phys.}\ }\textbf {\bibinfo {volume} {16}},\ \bibinfo {pages}
  {546} (\bibinfo {year} {2020})}\BibitemShut {NoStop}%
\bibitem [{\citenamefont {Por\'ee}\ \emph {et~al.}(2022)\citenamefont
  {Por\'ee}, \citenamefont {Lhotel}, \citenamefont {Petit}, \citenamefont
  {Krajewska}, \citenamefont {Puphal}, \citenamefont {Clark}, \citenamefont
  {Pomjakushin}, \citenamefont {Walker}, \citenamefont {Gauthier},
  \citenamefont {Gawryluk},\ and\ \citenamefont {Sibille}}]{Poree22}%
  \BibitemOpen
  \bibfield  {author} {\bibinfo {author} {\bibfnamefont {V.}~\bibnamefont
  {Por\'ee}}, \bibinfo {author} {\bibfnamefont {E.}~\bibnamefont {Lhotel}},
  \bibinfo {author} {\bibfnamefont {S.}~\bibnamefont {Petit}}, \bibinfo
  {author} {\bibfnamefont {A.}~\bibnamefont {Krajewska}}, \bibinfo {author}
  {\bibfnamefont {P.}~\bibnamefont {Puphal}}, \bibinfo {author} {\bibfnamefont
  {A.~H.}\ \bibnamefont {Clark}}, \bibinfo {author} {\bibfnamefont
  {V.}~\bibnamefont {Pomjakushin}}, \bibinfo {author} {\bibfnamefont {H.~C.}\
  \bibnamefont {Walker}}, \bibinfo {author} {\bibfnamefont {N.}~\bibnamefont
  {Gauthier}}, \bibinfo {author} {\bibfnamefont {D.~J.}\ \bibnamefont
  {Gawryluk}},\ and\ \bibinfo {author} {\bibfnamefont {R.}~\bibnamefont
  {Sibille}},\ }\bibfield  {title} {\bibinfo {title} {Crystal-field states and
  defect levels in candidate quantum spin ice
  ${\mathrm{ce}}_{2}{\mathrm{hf}}_{2}{\mathrm{o}}_{7}$},\ }\href
  {https://doi.org/10.1103/PhysRevMaterials.6.044406} {\bibfield  {journal}
  {\bibinfo  {journal} {Phys. Rev. Mater.}\ }\textbf {\bibinfo {volume} {6}},\
  \bibinfo {pages} {044406} (\bibinfo {year} {2022})}\BibitemShut {NoStop}%
\bibitem [{\citenamefont {Porée}\ \emph {et~al.}(2024)\citenamefont {Porée},
  \citenamefont {Bhardwaj}, \citenamefont {Lhotel}, \citenamefont {Petit},
  \citenamefont {Gauthier}, \citenamefont {Yan}, \citenamefont {Pomjakushin},
  \citenamefont {Ollivier}, \citenamefont {Quilliam}, \citenamefont
  {Nevidomskyy}, \citenamefont {Changlani},\ and\ \citenamefont
  {Sibille}}]{poree2024}%
  \BibitemOpen
  \bibfield  {author} {\bibinfo {author} {\bibfnamefont {V.}~\bibnamefont
  {Porée}}, \bibinfo {author} {\bibfnamefont {A.}~\bibnamefont {Bhardwaj}},
  \bibinfo {author} {\bibfnamefont {E.}~\bibnamefont {Lhotel}}, \bibinfo
  {author} {\bibfnamefont {S.}~\bibnamefont {Petit}}, \bibinfo {author}
  {\bibfnamefont {N.}~\bibnamefont {Gauthier}}, \bibinfo {author}
  {\bibfnamefont {H.}~\bibnamefont {Yan}}, \bibinfo {author} {\bibfnamefont
  {V.}~\bibnamefont {Pomjakushin}}, \bibinfo {author} {\bibfnamefont
  {J.}~\bibnamefont {Ollivier}}, \bibinfo {author} {\bibfnamefont {J.~A.}\
  \bibnamefont {Quilliam}}, \bibinfo {author} {\bibfnamefont {A.~H.}\
  \bibnamefont {Nevidomskyy}}, \bibinfo {author} {\bibfnamefont {H.~J.}\
  \bibnamefont {Changlani}},\ and\ \bibinfo {author} {\bibfnamefont
  {R.}~\bibnamefont {Sibille}},\ }\href@noop {} {\bibinfo {title}
  {Dipolar-octupolar correlations and hierarchy of exchange interactions in
  ce$_2$hf$_2$o$_7$}} (\bibinfo {year} {2024}),\ \Eprint
  {https://arxiv.org/abs/2305.08261} {arXiv:2305.08261 [cond-mat.str-el]}
  \BibitemShut {NoStop}%
\bibitem [{\citenamefont {Gao}\ \emph {et~al.}(2019)\citenamefont {Gao},
  \citenamefont {Chen}, \citenamefont {Tam}, \citenamefont {Huang},
  \citenamefont {Sasmal}, \citenamefont {Adroja}, \citenamefont {Ye},
  \citenamefont {Cao}, \citenamefont {Sala}, \citenamefont {Stone},
  \citenamefont {Baines}, \citenamefont {Verezhak}, \citenamefont {Hu},
  \citenamefont {Chung}, \citenamefont {Xu}, \citenamefont {Cheong},
  \citenamefont {Nallaiyan}, \citenamefont {Spagna}, \citenamefont {Maple},
  \citenamefont {Nevidomskyy}, \citenamefont {Morosan}, \citenamefont {Chen},\
  and\ \citenamefont {Dai}}]{Gao2019}%
  \BibitemOpen
  \bibfield  {author} {\bibinfo {author} {\bibfnamefont {B.}~\bibnamefont
  {Gao}}, \bibinfo {author} {\bibfnamefont {T.}~\bibnamefont {Chen}}, \bibinfo
  {author} {\bibfnamefont {D.~W.}\ \bibnamefont {Tam}}, \bibinfo {author}
  {\bibfnamefont {C.-L.}\ \bibnamefont {Huang}}, \bibinfo {author}
  {\bibfnamefont {K.}~\bibnamefont {Sasmal}}, \bibinfo {author} {\bibfnamefont
  {D.~T.}\ \bibnamefont {Adroja}}, \bibinfo {author} {\bibfnamefont
  {F.}~\bibnamefont {Ye}}, \bibinfo {author} {\bibfnamefont {H.}~\bibnamefont
  {Cao}}, \bibinfo {author} {\bibfnamefont {G.}~\bibnamefont {Sala}}, \bibinfo
  {author} {\bibfnamefont {M.~B.}\ \bibnamefont {Stone}}, \bibinfo {author}
  {\bibfnamefont {C.}~\bibnamefont {Baines}}, \bibinfo {author} {\bibfnamefont
  {J.~A.~T.}\ \bibnamefont {Verezhak}}, \bibinfo {author} {\bibfnamefont
  {H.}~\bibnamefont {Hu}}, \bibinfo {author} {\bibfnamefont {J.-H.}\
  \bibnamefont {Chung}}, \bibinfo {author} {\bibfnamefont {X.}~\bibnamefont
  {Xu}}, \bibinfo {author} {\bibfnamefont {S.-W.}\ \bibnamefont {Cheong}},
  \bibinfo {author} {\bibfnamefont {M.}~\bibnamefont {Nallaiyan}}, \bibinfo
  {author} {\bibfnamefont {S.}~\bibnamefont {Spagna}}, \bibinfo {author}
  {\bibfnamefont {M.~B.}\ \bibnamefont {Maple}}, \bibinfo {author}
  {\bibfnamefont {A.~H.}\ \bibnamefont {Nevidomskyy}}, \bibinfo {author}
  {\bibfnamefont {E.}~\bibnamefont {Morosan}}, \bibinfo {author} {\bibfnamefont
  {G.}~\bibnamefont {Chen}},\ and\ \bibinfo {author} {\bibfnamefont
  {P.}~\bibnamefont {Dai}},\ }\bibfield  {title} {\bibinfo {title}
  {Experimental signatures of a three-dimensional quantum spin liquid in
  effective spin-1/2 ce2zr2o7 pyrochlore},\ }\href
  {https://doi.org/10.1038/s41567-019-0577-6} {\bibfield  {journal} {\bibinfo
  {journal} {Nature Physics}\ }\textbf {\bibinfo {volume} {15}},\ \bibinfo
  {pages} {1052} (\bibinfo {year} {2019})}\BibitemShut {NoStop}%
\bibitem [{\citenamefont {Smith}\ \emph {et~al.}(2023)\citenamefont {Smith},
  \citenamefont {Dudemaine}, \citenamefont {Placke}, \citenamefont {Sch\"afer},
  \citenamefont {Yahne}, \citenamefont {DeLazzer}, \citenamefont {Fitterman},
  \citenamefont {Beare}, \citenamefont {Gaudet}, \citenamefont {Buhariwalla},
  \citenamefont {Podlesnyak}, \citenamefont {Xu}, \citenamefont {Clancy},
  \citenamefont {Movshovich}, \citenamefont {Luke}, \citenamefont {Ross},
  \citenamefont {Moessner}, \citenamefont {Benton}, \citenamefont {Bianchi},\
  and\ \citenamefont {Gaulin}}]{Smith2023}%
  \BibitemOpen
  \bibfield  {author} {\bibinfo {author} {\bibfnamefont {E.~M.}\ \bibnamefont
  {Smith}}, \bibinfo {author} {\bibfnamefont {J.}~\bibnamefont {Dudemaine}},
  \bibinfo {author} {\bibfnamefont {B.}~\bibnamefont {Placke}}, \bibinfo
  {author} {\bibfnamefont {R.}~\bibnamefont {Sch\"afer}}, \bibinfo {author}
  {\bibfnamefont {D.~R.}\ \bibnamefont {Yahne}}, \bibinfo {author}
  {\bibfnamefont {T.}~\bibnamefont {DeLazzer}}, \bibinfo {author}
  {\bibfnamefont {A.}~\bibnamefont {Fitterman}}, \bibinfo {author}
  {\bibfnamefont {J.}~\bibnamefont {Beare}}, \bibinfo {author} {\bibfnamefont
  {J.}~\bibnamefont {Gaudet}}, \bibinfo {author} {\bibfnamefont {C.~R.~C.}\
  \bibnamefont {Buhariwalla}}, \bibinfo {author} {\bibfnamefont
  {A.}~\bibnamefont {Podlesnyak}}, \bibinfo {author} {\bibfnamefont
  {G.}~\bibnamefont {Xu}}, \bibinfo {author} {\bibfnamefont {J.~P.}\
  \bibnamefont {Clancy}}, \bibinfo {author} {\bibfnamefont {R.}~\bibnamefont
  {Movshovich}}, \bibinfo {author} {\bibfnamefont {G.~M.}\ \bibnamefont
  {Luke}}, \bibinfo {author} {\bibfnamefont {K.~A.}\ \bibnamefont {Ross}},
  \bibinfo {author} {\bibfnamefont {R.}~\bibnamefont {Moessner}}, \bibinfo
  {author} {\bibfnamefont {O.}~\bibnamefont {Benton}}, \bibinfo {author}
  {\bibfnamefont {A.~D.}\ \bibnamefont {Bianchi}},\ and\ \bibinfo {author}
  {\bibfnamefont {B.~D.}\ \bibnamefont {Gaulin}},\ }\bibfield  {title}
  {\bibinfo {title} {Quantum spin ice response to a magnetic field in the
  dipole-octupole pyrochlore
  ${\mathrm{ce}}_{2}{\mathrm{zr}}_{2}{\mathrm{o}}_{7}$},\ }\href
  {https://doi.org/10.1103/PhysRevB.108.054438} {\bibfield  {journal} {\bibinfo
   {journal} {Phys. Rev. B}\ }\textbf {\bibinfo {volume} {108}},\ \bibinfo
  {pages} {054438} (\bibinfo {year} {2023})}\BibitemShut {NoStop}%
\bibitem [{\citenamefont {Yahne}\ \emph {et~al.}(2024)\citenamefont {Yahne},
  \citenamefont {Placke}, \citenamefont {Sch\"afer}, \citenamefont {Benton},
  \citenamefont {Moessner}, \citenamefont {Powell}, \citenamefont {Kolis},
  \citenamefont {Pasco}, \citenamefont {May}, \citenamefont {Frontzek},
  \citenamefont {Smith}, \citenamefont {Gaulin}, \citenamefont {Calder},\ and\
  \citenamefont {Ross}}]{Yahne2024}%
  \BibitemOpen
  \bibfield  {author} {\bibinfo {author} {\bibfnamefont {D.~R.}\ \bibnamefont
  {Yahne}}, \bibinfo {author} {\bibfnamefont {B.}~\bibnamefont {Placke}},
  \bibinfo {author} {\bibfnamefont {R.}~\bibnamefont {Sch\"afer}}, \bibinfo
  {author} {\bibfnamefont {O.}~\bibnamefont {Benton}}, \bibinfo {author}
  {\bibfnamefont {R.}~\bibnamefont {Moessner}}, \bibinfo {author}
  {\bibfnamefont {M.}~\bibnamefont {Powell}}, \bibinfo {author} {\bibfnamefont
  {J.~W.}\ \bibnamefont {Kolis}}, \bibinfo {author} {\bibfnamefont {C.~M.}\
  \bibnamefont {Pasco}}, \bibinfo {author} {\bibfnamefont {A.~F.}\ \bibnamefont
  {May}}, \bibinfo {author} {\bibfnamefont {M.~D.}\ \bibnamefont {Frontzek}},
  \bibinfo {author} {\bibfnamefont {E.~M.}\ \bibnamefont {Smith}}, \bibinfo
  {author} {\bibfnamefont {B.~D.}\ \bibnamefont {Gaulin}}, \bibinfo {author}
  {\bibfnamefont {S.}~\bibnamefont {Calder}},\ and\ \bibinfo {author}
  {\bibfnamefont {K.~A.}\ \bibnamefont {Ross}},\ }\bibfield  {title} {\bibinfo
  {title} {Dipolar spin ice regime proximate to an all-in-all-out n\'eel ground
  state in the dipolar-octupolar pyrochlore
  ${\mathrm{ce}}_{2}{\mathrm{sn}}_{2}{\mathrm{o}}_{7}$},\ }\href
  {https://doi.org/10.1103/PhysRevX.14.011005} {\bibfield  {journal} {\bibinfo
  {journal} {Phys. Rev. X}\ }\textbf {\bibinfo {volume} {14}},\ \bibinfo
  {pages} {011005} (\bibinfo {year} {2024})}\BibitemShut {NoStop}%
\bibitem [{\citenamefont {Lee}\ \emph {et~al.}(2012)\citenamefont {Lee},
  \citenamefont {Onoda},\ and\ \citenamefont {Balents}}]{Lee2012}%
  \BibitemOpen
  \bibfield  {author} {\bibinfo {author} {\bibfnamefont {S.}~\bibnamefont
  {Lee}}, \bibinfo {author} {\bibfnamefont {S.}~\bibnamefont {Onoda}},\ and\
  \bibinfo {author} {\bibfnamefont {L.}~\bibnamefont {Balents}},\ }\bibfield
  {title} {\bibinfo {title} {Generic quantum spin ice},\ }\href
  {https://doi.org/10.1103/PhysRevB.86.104412} {\bibfield  {journal} {\bibinfo
  {journal} {Phys. Rev. B}\ }\textbf {\bibinfo {volume} {86}},\ \bibinfo
  {pages} {104412} (\bibinfo {year} {2012})}\BibitemShut {NoStop}%
\bibitem [{\citenamefont {Desrochers}\ \emph {et~al.}(2023)\citenamefont
  {Desrochers}, \citenamefont {Chern},\ and\ \citenamefont
  {Kim}}]{Desrochers2023}%
  \BibitemOpen
  \bibfield  {author} {\bibinfo {author} {\bibfnamefont {F.}~\bibnamefont
  {Desrochers}}, \bibinfo {author} {\bibfnamefont {L.~E.}\ \bibnamefont
  {Chern}},\ and\ \bibinfo {author} {\bibfnamefont {Y.~B.}\ \bibnamefont
  {Kim}},\ }\bibfield  {title} {\bibinfo {title} {Symmetry fractionalization in
  the gauge mean-field theory of quantum spin ice},\ }\href
  {https://doi.org/10.1103/PhysRevB.107.064404} {\bibfield  {journal} {\bibinfo
   {journal} {Phys. Rev. B}\ }\textbf {\bibinfo {volume} {107}},\ \bibinfo
  {pages} {064404} (\bibinfo {year} {2023})}\BibitemShut {NoStop}%
\bibitem [{\citenamefont {Desrochers}\ and\ \citenamefont
  {Kim}(2024{\natexlab{a}})}]{Desrochers24}%
  \BibitemOpen
  \bibfield  {author} {\bibinfo {author} {\bibfnamefont {F.}~\bibnamefont
  {Desrochers}}\ and\ \bibinfo {author} {\bibfnamefont {Y.~B.}\ \bibnamefont
  {Kim}},\ }\bibfield  {title} {\bibinfo {title} {Spectroscopic signatures of
  fractionalization in octupolar quantum spin ice},\ }\href
  {https://doi.org/10.1103/PhysRevLett.132.066502} {\bibfield  {journal}
  {\bibinfo  {journal} {Phys. Rev. Lett.}\ }\textbf {\bibinfo {volume} {132}},\
  \bibinfo {pages} {066502} (\bibinfo {year} {2024}{\natexlab{a}})}\BibitemShut
  {NoStop}%
\bibitem [{\citenamefont {Rau}\ and\ \citenamefont {Gingras}(2019)}]{Rau19}%
  \BibitemOpen
  \bibfield  {author} {\bibinfo {author} {\bibfnamefont {J.~G.}\ \bibnamefont
  {Rau}}\ and\ \bibinfo {author} {\bibfnamefont {M.~J.}\ \bibnamefont
  {Gingras}},\ }\bibfield  {title} {\bibinfo {title} {Frustrated quantum
  rare-earth pyrochlores},\ }\href
  {https://doi.org/10.1146/annurev-conmatphys-022317-110520} {\bibfield
  {journal} {\bibinfo  {journal} {Annual Review of Condensed Matter Physics}\
  }\textbf {\bibinfo {volume} {10}},\ \bibinfo {pages} {357} (\bibinfo {year}
  {2019})},\ \Eprint
  {https://arxiv.org/abs/https://doi.org/10.1146/annurev-conmatphys-022317-110520}
  {https://doi.org/10.1146/annurev-conmatphys-022317-110520} \BibitemShut
  {NoStop}%
\bibitem [{\citenamefont {Smith}\ \emph {et~al.}(2022)\citenamefont {Smith},
  \citenamefont {Benton}, \citenamefont {Yahne}, \citenamefont {Placke},
  \citenamefont {Sch\"afer}, \citenamefont {Gaudet}, \citenamefont {Dudemaine},
  \citenamefont {Fitterman}, \citenamefont {Beare}, \citenamefont {Wildes},
  \citenamefont {Bhattacharya}, \citenamefont {DeLazzer}, \citenamefont
  {Buhariwalla}, \citenamefont {Butch}, \citenamefont {Movshovich},
  \citenamefont {Garrett}, \citenamefont {Marjerrison}, \citenamefont {Clancy},
  \citenamefont {Kermarrec}, \citenamefont {Luke}, \citenamefont {Bianchi},
  \citenamefont {Ross},\ and\ \citenamefont {Gaulin}}]{Smith22}%
  \BibitemOpen
  \bibfield  {author} {\bibinfo {author} {\bibfnamefont {E.~M.}\ \bibnamefont
  {Smith}}, \bibinfo {author} {\bibfnamefont {O.}~\bibnamefont {Benton}},
  \bibinfo {author} {\bibfnamefont {D.~R.}\ \bibnamefont {Yahne}}, \bibinfo
  {author} {\bibfnamefont {B.}~\bibnamefont {Placke}}, \bibinfo {author}
  {\bibfnamefont {R.}~\bibnamefont {Sch\"afer}}, \bibinfo {author}
  {\bibfnamefont {J.}~\bibnamefont {Gaudet}}, \bibinfo {author} {\bibfnamefont
  {J.}~\bibnamefont {Dudemaine}}, \bibinfo {author} {\bibfnamefont
  {A.}~\bibnamefont {Fitterman}}, \bibinfo {author} {\bibfnamefont
  {J.}~\bibnamefont {Beare}}, \bibinfo {author} {\bibfnamefont {A.~R.}\
  \bibnamefont {Wildes}}, \bibinfo {author} {\bibfnamefont {S.}~\bibnamefont
  {Bhattacharya}}, \bibinfo {author} {\bibfnamefont {T.}~\bibnamefont
  {DeLazzer}}, \bibinfo {author} {\bibfnamefont {C.~R.~C.}\ \bibnamefont
  {Buhariwalla}}, \bibinfo {author} {\bibfnamefont {N.~P.}\ \bibnamefont
  {Butch}}, \bibinfo {author} {\bibfnamefont {R.}~\bibnamefont {Movshovich}},
  \bibinfo {author} {\bibfnamefont {J.~D.}\ \bibnamefont {Garrett}}, \bibinfo
  {author} {\bibfnamefont {C.~A.}\ \bibnamefont {Marjerrison}}, \bibinfo
  {author} {\bibfnamefont {J.~P.}\ \bibnamefont {Clancy}}, \bibinfo {author}
  {\bibfnamefont {E.}~\bibnamefont {Kermarrec}}, \bibinfo {author}
  {\bibfnamefont {G.~M.}\ \bibnamefont {Luke}}, \bibinfo {author}
  {\bibfnamefont {A.~D.}\ \bibnamefont {Bianchi}}, \bibinfo {author}
  {\bibfnamefont {K.~A.}\ \bibnamefont {Ross}},\ and\ \bibinfo {author}
  {\bibfnamefont {B.~D.}\ \bibnamefont {Gaulin}},\ }\bibfield  {title}
  {\bibinfo {title} {Case for a ${\mathrm{u}(1)}_{\ensuremath{\pi}}$ quantum
  spin liquid ground state in the dipole-octupole pyrochlore
  ${\mathrm{ce}}_{2}{\mathrm{zr}}_{2}{\mathrm{o}}_{7}$},\ }\href
  {https://doi.org/10.1103/PhysRevX.12.021015} {\bibfield  {journal} {\bibinfo
  {journal} {Phys. Rev. X}\ }\textbf {\bibinfo {volume} {12}},\ \bibinfo
  {pages} {021015} (\bibinfo {year} {2022})}\BibitemShut {NoStop}%
\bibitem [{\citenamefont {Huang}\ \emph {et~al.}(2014)\citenamefont {Huang},
  \citenamefont {Chen},\ and\ \citenamefont {Hermele}}]{Huang14}%
  \BibitemOpen
  \bibfield  {author} {\bibinfo {author} {\bibfnamefont {Y.-P.}\ \bibnamefont
  {Huang}}, \bibinfo {author} {\bibfnamefont {G.}~\bibnamefont {Chen}},\ and\
  \bibinfo {author} {\bibfnamefont {M.}~\bibnamefont {Hermele}},\ }\bibfield
  {title} {\bibinfo {title} {Quantum spin ices and topological phases from
  dipolar-octupolar doublets on the pyrochlore lattice},\ }\href
  {https://doi.org/10.1103/PhysRevLett.112.167203} {\bibfield  {journal}
  {\bibinfo  {journal} {Phys. Rev. Lett.}\ }\textbf {\bibinfo {volume} {112}},\
  \bibinfo {pages} {167203} (\bibinfo {year} {2014})}\BibitemShut {NoStop}%
\bibitem [{\citenamefont {Udagawa}\ and\ \citenamefont
  {Moessner}(2019)}]{Udagawa19}%
  \BibitemOpen
  \bibfield  {author} {\bibinfo {author} {\bibfnamefont {M.}~\bibnamefont
  {Udagawa}}\ and\ \bibinfo {author} {\bibfnamefont {R.}~\bibnamefont
  {Moessner}},\ }\bibfield  {title} {\bibinfo {title} {Spectrum of itinerant
  fractional excitations in quantum spin ice},\ }\href
  {https://doi.org/10.1103/PhysRevLett.122.117201} {\bibfield  {journal}
  {\bibinfo  {journal} {Phys. Rev. Lett.}\ }\textbf {\bibinfo {volume} {122}},\
  \bibinfo {pages} {117201} (\bibinfo {year} {2019})}\BibitemShut {NoStop}%
\bibitem [{sup()}]{supp}%
  \BibitemOpen
  \href@noop {} {}\bibinfo {note} {See Supplemental Material at
  URL-will-be-inserted-by-publisher for details of exact diagonalization
  calculations, and some further mathematical details for Husimi cactus and
  GMFT calculations. Includes Refs. [38-39]}\BibitemShut {NoStop}%
\bibitem [{\citenamefont {Desrochers}\ and\ \citenamefont
  {Kim}(2024{\natexlab{b}})}]{desrochers2024finite}%
  \BibitemOpen
  \bibfield  {author} {\bibinfo {author} {\bibfnamefont {F.}~\bibnamefont
  {Desrochers}}\ and\ \bibinfo {author} {\bibfnamefont {Y.~B.}\ \bibnamefont
  {Kim}},\ }\href@noop {} {\bibinfo {title} {Finite temperature dynamics in
  0-flux and $\pi$-flux quantum spin ice: Self-consistent exclusive boson
  approach}} (\bibinfo {year} {2024}{\natexlab{b}}),\ \Eprint
  {https://arxiv.org/abs/2401.09551} {arXiv:2401.09551 [cond-mat.str-el]}
  \BibitemShut {NoStop}%
\bibitem [{\citenamefont {Wei}\ and\ \citenamefont {Curnoe}()}]{pyrochlore_ED}%
  \BibitemOpen
  \bibfield  {author} {\bibinfo {author} {\bibfnamefont {C.}~\bibnamefont
  {Wei}}\ and\ \bibinfo {author} {\bibfnamefont {S.~H.}\ \bibnamefont
  {Curnoe}},\ }\bibfield  {title} {\bibinfo {title} {Symmetry considerations in
  exact diagonalization: spin-1/2 pyrochlore magnets},\ }\bibinfo {note}
  {arXiv:2309.10670}\BibitemShut {NoStop}%
\bibitem [{\citenamefont {Hao}\ \emph {et~al.}(2014)\citenamefont {Hao},
  \citenamefont {Day},\ and\ \citenamefont {Gingras}}]{Hao14}%
  \BibitemOpen
  \bibfield  {author} {\bibinfo {author} {\bibfnamefont {Z.}~\bibnamefont
  {Hao}}, \bibinfo {author} {\bibfnamefont {A.~G.~R.}\ \bibnamefont {Day}},\
  and\ \bibinfo {author} {\bibfnamefont {M.~J.~P.}\ \bibnamefont {Gingras}},\
  }\bibfield  {title} {\bibinfo {title} {Bosonic many-body theory of quantum
  spin ice},\ }\href {https://doi.org/10.1103/PhysRevB.90.214430} {\bibfield
  {journal} {\bibinfo  {journal} {Phys. Rev. B}\ }\textbf {\bibinfo {volume}
  {90}},\ \bibinfo {pages} {214430} (\bibinfo {year} {2014})}\BibitemShut
  {NoStop}%
\end{thebibliography}%
\nocite{pyrochlore_ED}
\nocite{Hao14}

\end{document}